\documentclass[pra,showpacs,twocolumn,superscriptaddress,longbibliography]{revtex4-2}
\usepackage{amsmath}   
\usepackage{amssymb}
\usepackage{graphicx}
\usepackage{dcolumn}
\usepackage{bm}
\usepackage{amsmath}
\usepackage{graphicx}
\usepackage{amsfonts}
\usepackage{subfigure}
\usepackage{graphicx}
\usepackage{float}
\usepackage{color}
\usepackage{xcolor}
\usepackage{soul}
\usepackage[normalem]{ulem}
\usepackage{braket}
\usepackage{amssymb}
\usepackage{hyperref}
\usepackage{comment}
\usepackage{xr}
\usepackage[T1]{fontenc}

\usepackage{titlesec}
\hypersetup{
    colorlinks=true,
    citecolor=black,
    linkcolor=blue,
    filecolor=black,      
    urlcolor=blue,
}
\begin{document}

\title{Trapped atoms and superradiance on an integrated nanophotonic microring circuit}

\author{Xinchao Zhou}
\affiliation{
 Department of Physics and Astronomy, Purdue University, West Lafayette, IN 47907, USA
}
\author{Hikaru Tamura}
\affiliation{
 Department of Physics and Astronomy, Purdue University, West Lafayette, IN 47907, USA
}
\author{Tzu-Han Chang}
\affiliation{
 Department of Physics and Astronomy, Purdue University, West Lafayette, IN 47907, USA
}
\author{Chen-Lung Hung}
\affiliation{
 Department of Physics and Astronomy, Purdue University, West Lafayette, IN 47907, USA
}
\affiliation{
 Purdue Quantum Science and Engineering Institute, Purdue University, West Lafayette, IN 47907, USA
}
\email{clhung@purdue.edu}

\date{\today}
\begin{abstract}
Interfacing cold atoms with integrated nanophotonic devices could offer new paradigms for engineering atom-light interactions and provide a potentially scalable route for quantum sensing, metrology, and quantum information processing. However, it remains a challenging task to efficiently trap a large ensemble of cold atoms on an integrated nanophotonic circuit. Here, we demonstrate direct loading of an ensemble of up to 70 atoms into an optical microtrap on a nanophotonic microring circuit. Efficient trap loading is achieved by employing degenerate Raman-sideband cooling in the microtrap, where a built-in spin-motion coupling arises directly from the vector light shift of the evanescent field potential on a microring. Atoms are cooled into the trap via optical pumping with a single free space beam. We have achieved a trap lifetime approaching 700\,ms under continuous cooling. We show that the trapped atoms display large cooperative coupling and superradiant decay into a whispering-gallery mode of the microring resonator, holding promise for explorations of new collective effects. Our technique can be extended to trapping a large ensemble of cold atoms on nanophotonic circuits for various quantum applications.
\end{abstract}

\maketitle

Integrating cold atomic ensemble with nanoscale dielectric structures \cite{2018RMP_DEChang} promises new and potentially scalable quantum applications from quantum nonlinear optics \cite{2014NaturePhotonics_chang}, quantum network \cite{2008Nature_KimbleInternet}, to sensing and metrology \cite{2019PRA_metrology}. Utilizing strong atom-light interactions, large photon-photon interaction can be engineered at a few-photon level \cite{2016PRX_PhotonMolecules, 2014Science_Dayan,2022NaturePhysics_QD, 2023NaturePhysics_photonboundstate, 2023RMP_wQED}. By coupling multiple atoms to a common photonic channel, new regimes of collective excitation and radiative dynamics can be accessed \cite{2015PRL_pcw, 2017NC_superradiance, 2022PRL_collecticedecay, 2023PRL_SuperradiancewQED, 2020PRL_Kanu, 2023PRL_ZenoRegime}. Similarly, photon-mediated long-range interaction between atoms can be engineered for simulating coherent \cite{2015NaturePhotonics_manybodymodels, 2015NaturePhotonics_subwavelength, 2016PNAS_hung, 2016PNAS_Hood_badedge, 2019ScienceAdvances_topologicalwQED, 2021Nature_ProgramableInteraction_monika} and dissipative \cite{2014PRL_QuantumSpinDimers} many-body quantum spin dynamics. Collective effects can also be utilized for photon storage and multiphoton generation with improved fidelity \cite{2017PRX_ana_photonstorage,2017PRL_multiphotongeneration}. Entangled states of atoms can be prepared \cite{2014Science_entangledStateFiberCavity, 2015Science_entanglementgeneration, 2019Nature_Laurat, 2018PRL_entanglement2atoms} and transported \cite{2022Sciecn_entanglementtransport, 2021NaturePhotonics_Rempe} via a coupled photonic structure and fiber network for quantum communication. For quantum sensing, trapped atomic ensemble along a nanophotonic waveguide may be spin-squeezed \cite{2018PRX_squeezingNanofiber, 2010PRL_CavitySqueezing} and detected with high efficiency, thus may provide metrology gain for atom interferometers \cite{2022NC_Lee, 2022APL_Yuri} and atomic clocks \cite{2019Optica_photonicIntegrationClock, 2020Nature_EntanglementClock, 2016PRL_networkofClock}. In quantum chemistry, ultracold molecules in a target rovibrational state may be directly photoassociated from cold atoms and interrogated through a strongly-coupled radiative channel with enhanced photonic density of states \cite{2017NJP_molecule, 2018NJP_molecules,2020PRL_MoleculeFormation}.

Realizing ensemble atom trapping on an integrated nanophotonic circuit would enable multiple quantum functionalities packed into one single optical chip. However, standard magneto-optical trapping (MOT) techniques have thus far only succeeded in loading atoms on suspended, one-dimensional nanostructures because of the multi-beam requirement to balance the radiation pressure during laser cooling. A few thousand atoms were cooled into a two-color evanescent field trap on nanofibers \cite{2010PRL_nanofiber, 2012PRL_nanofiber, 2023PRXQuantum_Srnanofiber}. Few-atom trapping was realized by overlapping a large-volume optical trap or by moving free space-loaded optical tweezers onto a suspended photonic crystal waveguide \cite{2015PRL_pcw} or cavity \cite{2013Science_tweezertrap}. For loading atoms onto integrated photonic circuits with a planar surface and with limited optical access from free space, optical conveyor belts \cite{2019PNAS_Kimble, 2019NC_imaging, 2023CPL_conveyor-belt} and an optical guiding method \cite{2023PRL_Coupling} may be implemented. Laser-cooled atoms have been transported to the near-field region for coupling to a nanophotonic microring resonator \cite{2023PRL_Coupling} or to a photonic crystal waveguide \cite{2019PNAS_Kimble}. 
Despite these prior efforts, however, direct laser-cooling atoms into near-field optical traps in the vicinity of 2D nanophotonic structures and stably trapping $N\gg1$ atoms with large cooperative atom-photon coupling on an integrated circuit have never been demonstrated.

Here, we report the first realization of efficient atom loading and trapping on an integrated nanophotonic circuit. We show that light-atom interaction in nanophotonics leads to a novel spin-motion coupling that directly enables degenerate Raman-sideband cooling (dRSC) \cite{1998PRL_dRSC, 2000PRL_dRSC}, thus providing a new route to directly bind and cool atoms to a near field trap. We demonstrate a large collective atom-photon coupling and a superradiant decay into a whispering-gallery-mode (WGM) resonator. Specifically, our method combines optical guiding (OG)~\cite{2023PRL_Coupling} and dRSC to efficiently load $N\approx 70$ cesium atoms into a small-volume $\sim O(10\mu$m$^3)$ optical microtrap formed on top of a microring resonator, achieving a trap lifetime approaching one second. 

\begin{figure}[!t]
\centering
\includegraphics[width=0.48\textwidth]{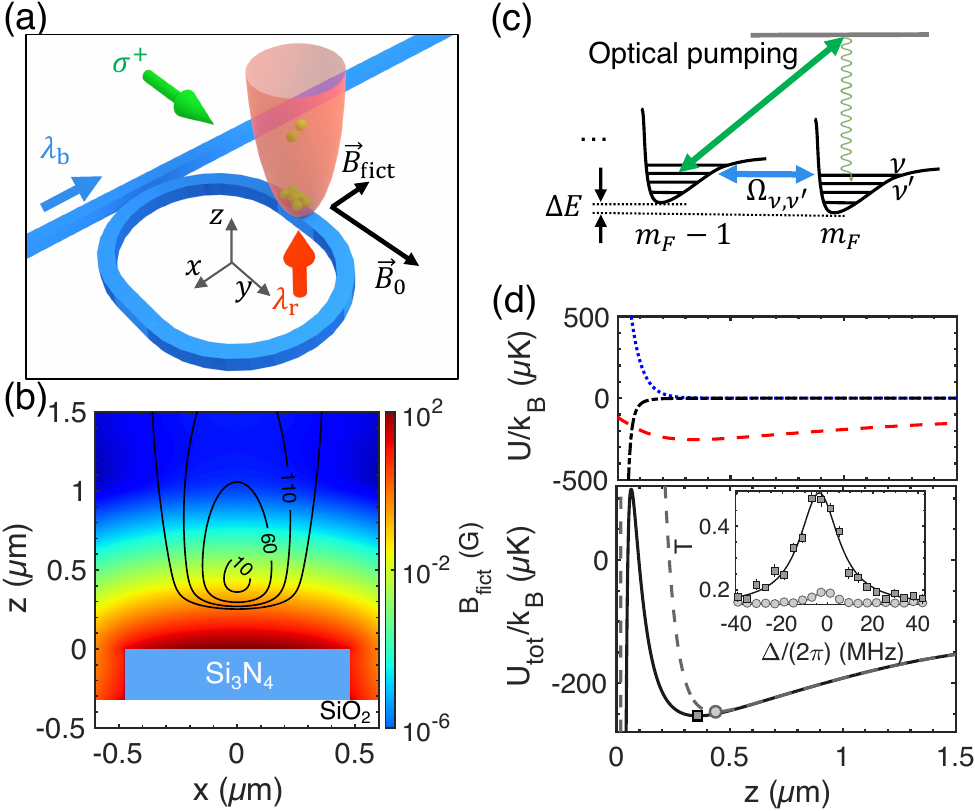}
\caption{Atom trapping on a nanophotonic microring circuit.
(a), Schematics of the experiment setup. Cold atoms (green spheres) are localized in an optical microtrap (red shade region) formed by a bottom-illuminating optical guiding beam (red arrow) and the evanescent field of a WGM excited from a bus waveguide (blue arrow). The latter also creates a fictitious magnetic field $\Vec{B}_{\mathrm{fict}}$ normal to the bias field $\Vec{B}_{0}$, which defines the quantization axis. 
Cooling and trap loading is realized by optical pumping (green arrow) for degenerate Raman-sideband cooling (dRSC).   
(b), Fictitious magnetic field above the microring waveguide. Overlaid contours mark the equipotential lines relative to the trap minimum, labeled in units of microKelvin. (c), Schematics of the dRSC. The fictitious field creates spin-motion coupling between degenerate trap states. The bias field adjusts the Zeeman splitting $\Delta E$. Optical pumping followed by spontaneous decay in the Lamb-Dicke regime cools the atomic motion. (d), Top panel: Linecuts of scalar potentials $U(0,0,z)$ of the optical guiding (red dashed curve), the repulsive barrier (blue dotted curve), and the Casimir-Polder interaction (black dashed-dotted curve), respectively; $k_\mathrm{B}$ is the Boltzmann constant. Bottom panel: Total scalar potential $U_{\mathrm{tot}}(0,0,z)$ with strong (dashed curve) and weak (solid curve) repulsive barrier during dRSC and probe stages, respectively. Filled symbols mark the location of trap centers. Inset compares probe transmission signals versus laser detuning $\Delta$ using these two trap configurations with the same number of trapped atoms. Error bars represent the standard error of the mean.}
\label{fig:fig1}
\end{figure}

Our approach connects OG to cooling and trapping in an evanescent field region. 
The microtrap is formed from a `red-color' bottom-illuminating beam with $10~\mu$m waist and 
wavelength $\lambda_{\mathrm{r}}\approx935\,\mathrm{nm}$, propagating along $\hat{z}$ and transmitting through the silica substrate of the circuit to form a funnel-like attractive potential (see Fig.~\ref{fig:fig1}). The funnel can guide a large flux of laser-cooled atoms onto a silicon-nitride microring resonator \cite{2023PRL_Coupling}. In the meantime, a `blue-color' WGM resonant at $\lambda_{\mathrm{b}}\approx849.552\,\mathrm{nm}$ is excited to create a repulsive evanescent-field potential barrier to plug the optical funnel. This prevents guided atoms from hitting the microring surface and forms a stable trap minimum within the guiding potential at a tunable distance $z_\mathrm{c}\approx 360 - 440~$nm above the microring.

The evanescent field potential also mediates spin-motion coupling, which is crucial for trap loading. For a WGM with transverse-field polarization perpendicular to the waveguide top surface, the evanescent field is $\sim 98\%$ circularly polarized above the waveguide due to a rapid decay in the transverse field~\cite{2016Optica_evanescent}. The polarization rotates about the $x$-axis, which is normal to the local mode propagation direction defined along $\hat{y}$. To the atoms, the circularly-polarized evanescent field imposes a position-dependent vector light shift equivalent to a fictitious magnetic field $\Vec{B}_{\mathrm{fict}} \propto -|\Vec{E}(\vec{r})|^2\hat{x}$ 
acting on the atoms, where $\Vec{E}$ is the complex mode field (see Appendix B). Significant spin-motion coupling arises because the fictitious field amplitude decays exponentially along $\hat{z}$, as shown in Fig.~\ref{fig:fig1}(b). The field varies more weakly and symmetrically along $\hat{x}$ and remains constant along $\hat{y}$. 
With large spin-motion coupling primarily along the $z$-axis, optically guided atoms can be Raman-cooled and bound to the microtrap by optical pumping \cite{2000PRL_dRSC}. We note that, when desired, this fictitious field can also be eliminated by injecting an equal-amplitude, counter-propagating mode~\cite{2010PRL_nanofiber, 2012PRL_nanofiber,2019Optica_Chang} with opposite fictitious field contribution. 

The actual cooling scheme works as follows. As illustrated in Fig.~\ref{fig:fig1}(a), we apply a bias magnetic field $\Vec{B}_{0}$ along $\hat{y}$ to define the quantization axis. The fictitious field creates a spin-motion Raman coupling rate $ \Omega_{\nu,\nu'} \sim 2\pi \times 10~$kHz between degenerate trap states of adjacent magnetic levels as in Fig.~\ref{fig:fig1}(c) (see Appendix C and Fig.~\ref{fig:fig6}); $\nu'<\nu$ denote trap vibrational levels along the $z$-axis 
and the trap potential $U_{\mathrm{tot}}(\Vec{r})$ has contributions from the total scalar light shifts of the OG beam and the blue-color evanescent field, together with an atom-surface Casimir-Polder potential that weakens the trap for $z\lesssim 100~$nm (Appendix F); see Fig.~\ref{fig:fig1}(d). 
Applying optical pumping with $\sigma^+$ transitions in the Lamb-Dicke regime, that is, light scattering without changing trap states allows trapped atoms to be pumped into a dark state with reduced energy. 

Our experiment begins with pre-cooling cesium atoms in the optical funnel using a MOT at $z\approx 300~\mu$m far above the circuit. The atoms are then polarized in the $F=3$ hyperfine ground state and are guided toward the surface microtrap. During this time, the blue-color evanescent field is kept fully on, supporting a stable trap center at $z_\mathrm{c}\approx 440~$nm. We perform dRSC to cool guided atoms near the surface, where an optical pumping beam drives primarily $\sigma^{+}$ transitions to the $|6P_{3/2},F^{\prime}=2 \rangle$ excited states. After $40\,$ms of cooling, atoms are fully polarized in the $\ket{F=3, m_{F}=3}$ state. We then pump them back to the $|F=4,m_F=4\rangle$ state and adjust $\vec{B}_0$ to reorient the quantization axis transverse to the waveguide (along $-\hat{x}$) for probing.

\begin{figure}[t]
\centering
\includegraphics[width=0.5\textwidth]{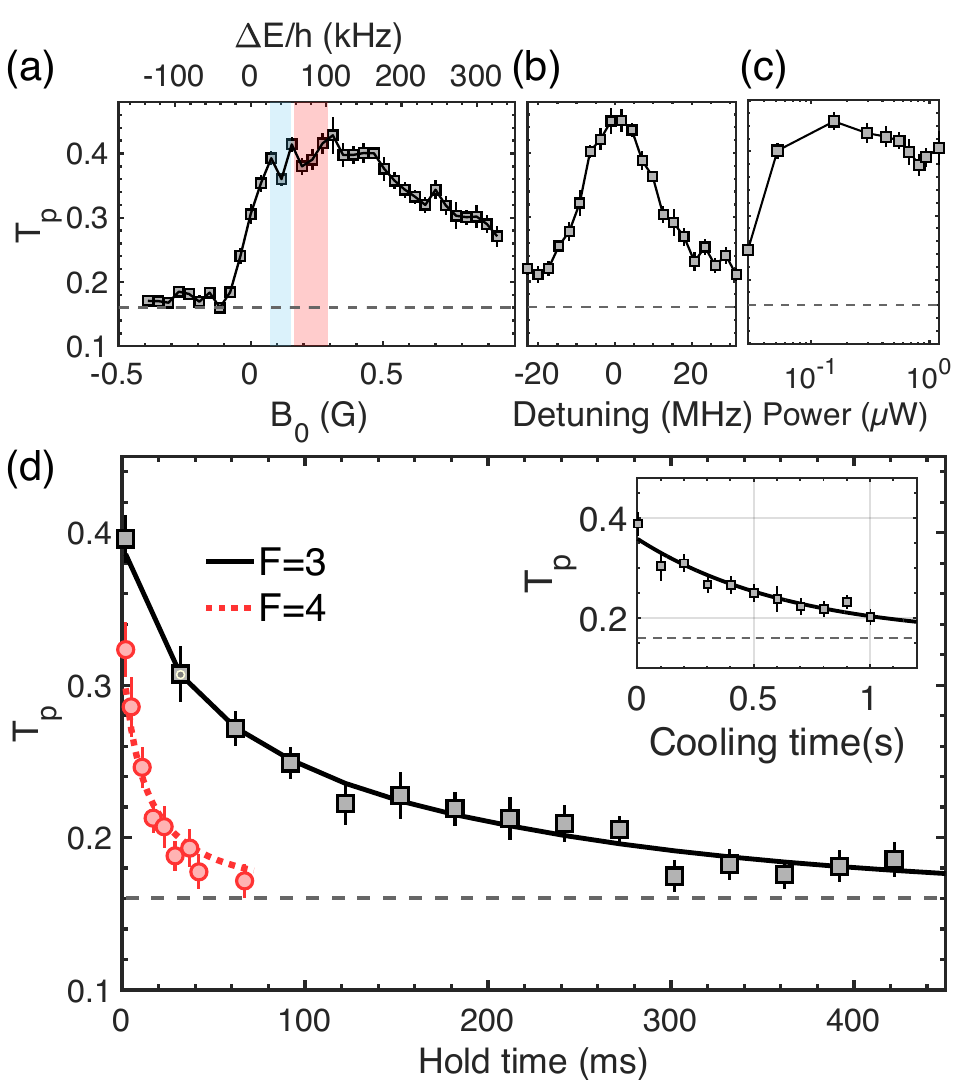}
\caption{Optimization of cooling parameters and lifetime of trapped atoms.
(a), Peak transmission $T_\mathrm{p}$ as a function of the bias magnetic field and the corresponding Zeeman splitting $\Delta E$. Negative $B_0$ indicates the direction is along $-\hat{y}$. Blue (red) shaded region marks the range of trap level spacing along the $z$-($x$-)axis. (b-c), Dependencies of $T_\mathrm{p}$ on the pump detuning relative to the $F=3 \leftrightarrow F^{\prime}=2$ resonance in free space (b) and the pump power (c). (d), Transmission versus hold time (black squares). Our fit (solid curve) indicates a one-body lifetime of $\tau\approx 230\,\mathrm{ms}$. For comparison, transmission with atoms repumped to the $F=4$ state after cooling (red circles) shows a faster decay well-fitted by a two-body loss model (dotted curve); see Appendix D. Inset shows measurement under continuous cooling, giving a $1/e$ lifetime of $\tau\approx 690\,\mathrm{ms}$. 
Dashed lines mark the transmission without atoms. Error bars represent the standard error of the mean.
}
\label{fig:fig2}
\end{figure}

Trapped atoms are detected via an atom-induced transparency \cite{2023PRL_Coupling}. Through the bus waveguide, we excite a `probe' WGM resonant with the $|F=4,m_F=4\rangle$ to $|F^{\prime}=5,m_{F^\prime}=5\rangle$ cycling transition at $\lambda_\mathrm{D2} \approx 852.352~$nm; see Fig.~\ref{fig:fig1}(d) inset. The presence of trapped atoms allows photons to transmit through the bus waveguide instead of being dissipated in the microring. We note that atom-WGM photon coupling strengths can be adjusted by changing the position of the trap center. This can be achieved by ramping down the power of the blue-color evanescent field shortly before probing. As shown in Fig.~\ref{fig:fig1}(d), the trap center can be adiabatically shifted towards $z_\mathrm{c}\approx 360\,$nm, and we indeed observe a significant increase in the waveguide transmission. In the following studies, we probe trapped atoms in this configuration.

\begin{figure*}[ht]
\centering
\includegraphics[width=1.0\textwidth]{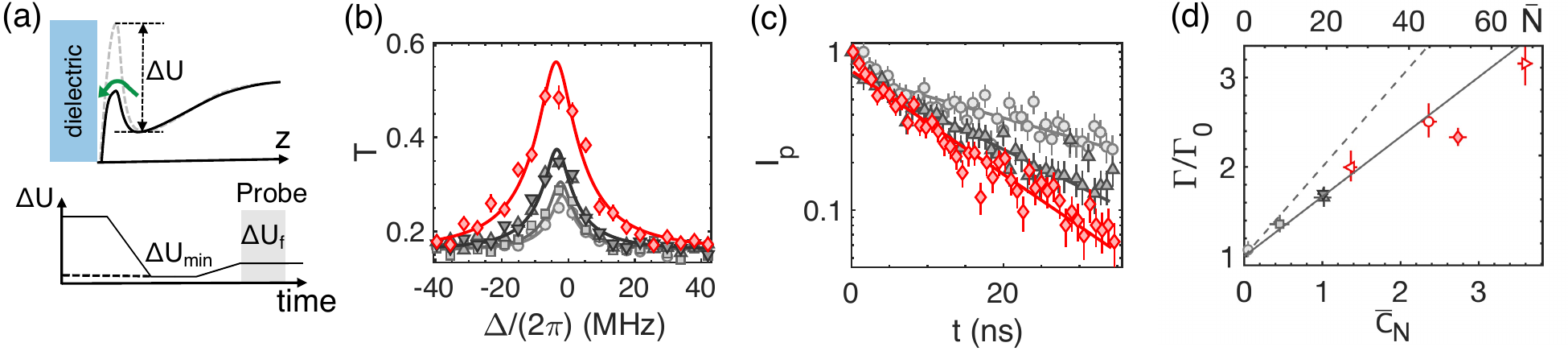}
\caption{Observation of cooperative coupling (b) and superradiant decay (c-d).
(a), Trap-spilling procedure. The number of atoms in the trap can be controlled by ramping down the repulsive barrier to different $\Delta U_{\mathrm{min}}$, followed by ramping back to the same final height $\Delta U_f$ for probing.
(b), Steady-state transmission spectra measured after several trap-spilling with $\Delta U_{\mathrm{min}}\approx k_{\mathrm{B}} \times 60$ (circles), 150 (squares), 210 (up-triangles), and 1020 (down-triangle)~$\mu$K, respectively. Additional spectrum (red diamonds) is measured with a larger initial atom number. 
Solid lines are theoretical fits (Appendix G) to extract the averaged cooperativity $\bar{C}_N$. 
(c), Normalized transmission counts $I_{p}$ of pulsed-excitation decay rate measurements (filled symbols), using identical loading procedures as in (b). 
The probe pulse is extinguished at $t=0\,\mathrm{ns}$ (see Appendix H and Fig.~\ref{fig:fig8}). Solid lines are exponential fits to extract the decay rate $\Gamma$. In (b-c), error bars are the standard error of the mean.
(d), Normalized decay rate $\Gamma/\Gamma_{0}$ versus cooperativity $\bar{C}_N$ measured in (b-c) (filled symbols). Additional measurements with larger initial atom numbers are plotted with open symbols. Top axis indicates mean atom number $\bar{N}=\bar{C}_N/\bar{C}_1$. 
The dashed line assumes $\Gamma = (1+\bar{C}_N)\Gamma_{0}$, and the solid line gives a reduced slope $\eta$ after considering the back-scattering of the probe WGM; see text. Error bars are the uncertainties of the fitted parameters.}
\label{fig:fig3}
\end{figure*}

\begin{figure}[b]
\centering
\includegraphics[width=0.49\textwidth]{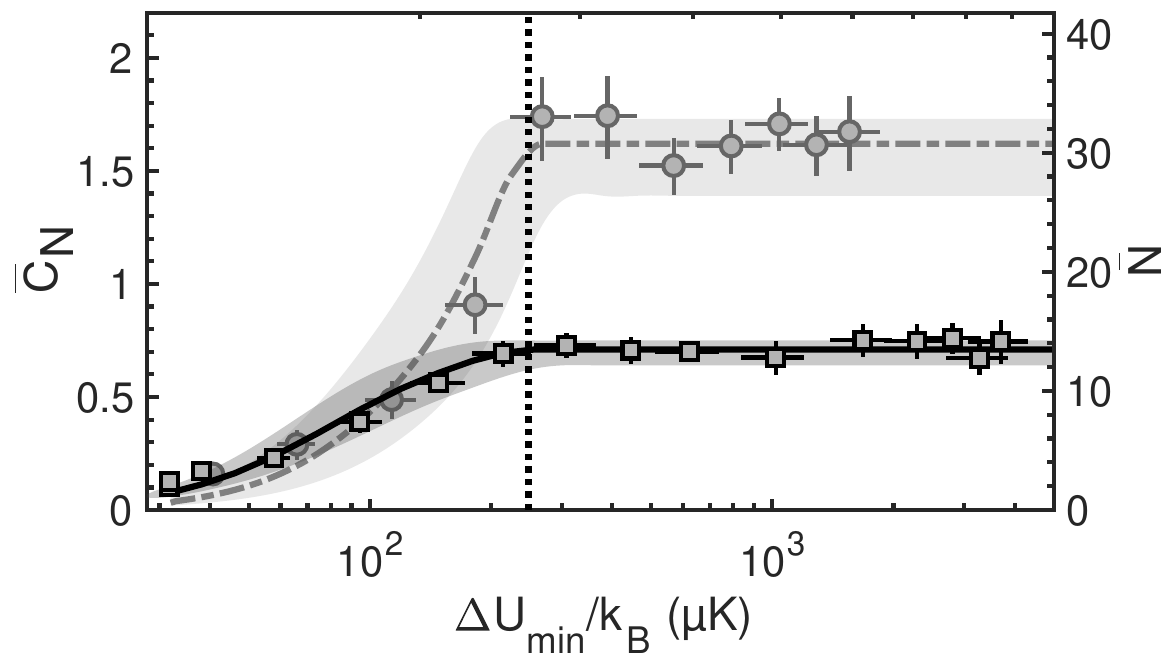}
\caption{Determination of temperature and atom number by controlled trap-spilling with different initial atom numbers.
Cooperativity $\bar{C}_{N}$ extracted from the resonant transmission is plotted as a function of $\Delta U_{\mathrm{min}}$. Solid and dashed curves are theoretical fits (see Appendix I), giving low temperatures $T_\mathrm{trap}=23(1)\,\mu$K and $38(5)\,\mu$K, respectively. Shaded bands include fit uncertainty. The vertical dotted line indicates the expected barrier height when spilling becomes effective. The right axis indicates the corresponding atom number $\bar{N}$. Error bars are the standard error of the mean.}
\label{fig:fig4}
\end{figure}

We first optimize the cooling performance by maximizing the peak transparency signal. Dependencies on the bias magnetic field magnitude $B_{0}$ and optical pumping are shown in Fig.~\ref{fig:fig2}(a-c). The optimal magnetic field leads to a Zeeman splitting near the level spacing $\Delta E  \sim \Delta E_\nu \sim h\times 30- 50~$kHz for $\nu<40$, where $h$ is the Planck constant. But the bias field dependence is quite broad, likely due to the large Raman coupling that could broaden trap levels and could also drive transitions for $|\nu-\nu^\prime| \geq 2$ (see Appendix C and Fig.~\ref{fig:fig6}). We further optimize $B_0\approx 150~$mG by minimizing the temperature discussed in Fig.~\ref{fig:fig4}. On the other hand, atoms are significantly heated out of the trap if the field is aligned to the opposite orientation, making the optical pumping beam drive $\sigma^-$ transitions. The optical pumping frequency is optimized to be resonant with the $F=3\leftrightarrow F^{\prime}=2$ transition, and the optimal power is $\approx 0.2~\mu$W, giving an intensity $\approx 150\,\mu\mathrm{W/cm}^2$. 

We measure long and state-dependent lifetimes in the trap following cooling. As shown in Fig.~\ref{fig:fig2}(d), we find a one-body lifetime $\approx 230~$ms for trapped atoms polarized in the $\ket{F=3,m_F=3}$ state. The transmission signal decays faster if we repump the atoms to the $F=4$ state immediately after cooling. This is consistent with two-body inelastic collision loss at a high density $\sim10^{13}/$cm$^{3}$ when cesium atoms are not polarized in the lowest-energy ground state (see Appendix D). We can extend the trap lifetime to nearly one second when we perform cooling continuously, as shown in the inset of Fig.~\ref{fig:fig2}(d). This is only about $2$ times shorter than the vacuum-limited lifetime measured from a free space MOT.

Remarkably, we observe cooperative atom-photon coupling through a variable number of stably trapped atoms on the circuit. This also allows us to deduce the number of atoms coupled to the microring. As illustrated in Fig.~\ref{fig:fig3}(a), we can reduce the number of trapped atoms while maintaining the same initial loading condition using a controlled trap-spilling procedure. Here, we ramp down the trap barrier height in 0.5~ms to a variable minimum value $\Delta U_\mathrm{min}$. We then hold for an additional $0.5~$ms, allowing energetic atoms to escape the trap, 
followed by ramping the barrier back to the same $\Delta U_f$ for probing the remaining atoms; $\Delta U_f$ is the minimum barrier height that we can ramp down to without losing atomic signal. 
In evaluating $\Delta U_\mathrm{min}$, we calculate the full potential including the vector light shift for the $|F=3, m_F=3\rangle$ state (Appendix B). 

Figure~\ref{fig:fig3}(b) shows the measured transmission spectra, where all the spectral linewidths are broader than the free space value $\Gamma_{0}=2\pi \times 5.2~\mathrm{MHz}$. The spectra with larger $\Delta U_{\mathrm{min}}$ or more atoms in the trap show more significant transparency and broader widths. This dependence suggests that trapped atoms are superradiantly coupled. 

We perform theoretical fits to the measured spectra, assuming that the atoms couple to the probe WGM with a collective cooperativity $C_N \equiv NC_{1}$, where $N$ is the effective atom number, $C_{1}=4g^2/(\kappa\Gamma^\prime)$ is the single-atom cooperativity, $g$ is the atom-WGM photon coupling rate, $\kappa$ is the decay rate of the WGM, and $\Gamma^\prime \approx \Gamma_0$ is the single atom decay rate to all other non-guided modes; see Appendix G. Our simple model neglects possible collective effects due to atoms coupled via the non-guided modes. We also note that trapped atoms will only couple to the probe WGM ($\sigma^+$ polarized) but nearly not to its counter-propagating mode ($\sigma^-$ polarized) because of a large asymmetry in the atom-photon coupling rate~\cite{2023PRL_Coupling,2017Nature_ChiralQuantumOptics}. 
The fitted spectra and fitted mean value $\bar{C}_N$ are displayed in Fig.~\ref{fig:fig3}(b) and (d), respectively.

We further confirm the cooperative coupling by observing superradiant decay from the excited state, as shown in Fig.~\ref{fig:fig3}(c). We excite the system with a weak short pulse (full width at half maximum  $\sim 6\,$ns which is much shorter than the natural excited-state lifetime) and extract the total decay rate by an exponential fit to the monitored transmitted counts after the excitation pulse (see Appendix H).  Our measurement indicates a significant speed-up for decaying into the WGM compared to the free space decay rate. In Fig.~\ref{fig:fig3}(d), we plot the measured total decay rate versus the fitted cooperativity $\bar{C}_N$ from the steady-state transmission spectrum, which shows a good agreement with the expectation $\Gamma/\Gamma_{0}\approx (1+\eta \bar{C}_N)$, where the reduction factor $\eta \approx 0.67 <1$ is due to probe photons being back-scattered to the counter-propagating WGM that is nearly uncoupled from the atoms (Appendix G). 

The trap-spilling procedure in Fig.~\ref{fig:fig3} can also be used to characterize the temperature and, subsequently, the number of trapped atoms. Figure~\ref{fig:fig4} shows two different examples of $\bar{C}_N$ versus barrier height $\Delta U_\mathrm{min}$. Trap spilling is only effective when the barrier becomes lower than the initial trap depth $|U_\mathrm{tot}(0,0,z_\mathrm{c})|\approx 250~\mu$K. Indeed, the measured $\bar{C}_N$ remains constant until $\Delta U_\mathrm{min}$ approaches this value, indicating our understanding of the trap, including the Casimir-Polder surface interaction, is sufficiently accurate. Once $\bar{C}_N$ is reduced after the trap spill, we determine the temperature by fitting a truncated Boltzmann distribution to evaluate the survival probability (Appendix I). Our fit gives a low temperature $T_\mathrm{trap}\approx 23~\mu\mathrm{K}$ and an averaged vibrational quantum number $\bar{\nu}\approx 14$. We estimate $\bar{\nu}_{x,y}\approx(5,36)$ along the $x$- and $y$-axis, respectively. The fit allows us to evaluate in-trap density distributions and estimate the single-atom cooperativity $\bar{C}_1\approx 0.05$ (for $z_\mathrm{c}\approx360~$nm). 
Subsequently, we determine the mean atom number $\bar{N}=\bar{C}_N/\bar{C}_1$ as shown in the right (top) axis of Fig.~\ref{fig:fig4} (Fig.~\ref{fig:fig3}(d)). The highest $\bar{C}_N\approx 3.6$ in Fig.~\ref{fig:fig3}(d) indicates we effectively localize $N\approx 68$ trapped atoms in a small microtrap.

When a dense ensemble of atoms are trapped in the surface microtrap and superradiantly excited by the probe WGM, additional collective effects due to coupling between atoms via the non-guided radiation modes may also manifest. For example, a collective Lamb shift~\cite{2016PRL_Browaeys_shifts,2016PRA_Francis_atomicarray, 2009PRL_Scully} can appear  as a weak dependence of resonance shift on the atom number $N$. Since the wavenumber of WGM $k_{\rm WGM} > 2\pi/\lambda_{\rm D2}$, these atoms may not couple well to the radiation modes in free space due to significant phase mismatch, leading to a collectively suppressed decay rate into free space~\cite{2017PRX_ana_photonstorage, 2016PRA_Francis_atomicarray}. These novel effects could be uniquely explored with densely trapped atoms on a microring circuit.

We expect that cooling and trapping performance could be further improved with minor adjustments. The microtrap can be expanded to cover the entire microring (circumference $\approx 150~\mu$m) to increase the trap volume and, thus, the number of trapped atoms by at least $10$ times. In this work, in-trap atomic density has reached $\sim 10^{13}$/cm$^{3}$. Light-assisted collisions and photon re-absorption during optical pumping could limit cooling performance. By using a tightly confining lattice for localizing individual atoms (such as a two-color evanescent field trap~\cite{2019Optica_Chang}), further cooling in all trap axes to near the vibrational ground states can be realized as pioneered in nanofibers~\cite{2018PRX_Nanofibercooling}. Meanwhile, inelastic collision loss can also be entirely suppressed. Transferring atoms to a tighter trap near the surface to $z_\mathrm{c}\approx 100~$nm can further increase single-atom cooperativity $C_1\gtrsim 10$ \cite{2019Optica_Chang,2023PRL_Coupling}, boosting $C_N > 1000$. We believe our platform and methodology could enable a large ensemble ($N\gtrsim 10^3$) of collectively-coupled atoms trapped on an integrated nanophotonic circuit. In the future, it would also be interesting to perform direct dRSC of trapped atoms into quantum degeneracy \cite{2017Science_Vladan, 2019PRL_Rb_Vladan, 2019PRL_Cesium_Vladan} or explore cavity-assisted molecule formation with the current platform \cite{2017NJP_molecule, 2018NJP_molecules,2020PRL_MoleculeFormation}. This would pave the road towards various applications in quantum nonlinear optics, quantum simulations, sensing, and quantum chemistry using cold atom-powered nanophotonic circuits. 

\section*{Acknowledgement}
We thank Francis Robicheaux, Deepak Suresh, and Jonathan Hood for discussions. We are grateful to Ming Zhu, Yuanhao Liang, and William Gomez for their prior assistance in the experimental work. This work was supported by the AFOSR (Grant NO. FA9550-22-1-0031) and the NSF (Grant NO. PHY-1848316 and ECCS-2134931).

\section*{Appendix A: Experiment setup}
Our nanophotonic circuit contains silicon-nitride microring resonators fabricated on a transparent SiO$_2$-Si$_3$N$_4$ double-layer
membrane, which is suspended over a 2 mm $\times$ 8 mm window on a silicon chip. A bus waveguide near-critically couples to a microring  (coupling rate $\kappa_\mathrm{e}$ comparable to the intrinsic resonator loss rate $\kappa_\mathrm{i}$) and is end-coupled to two cleaved fibers
for input and output with at least one port approaching $\gtrsim 80\%$ single-pass coupling efficiency~\cite{2022OE_3Dcoupler}. 

Cold atoms are introduced above the chip and cooled into the optical guiding potential using standard MOT and polarization-gradient cooling techniques. The optical guiding beam has a small waist of $10~\mu$m, focused on the membrane, and has a power of $21~$mW after transmitting through the membrane. The wavelength is tuned to $\lambda_r\approx 935~$nm to eliminate the differential light shift in the cesium D2 line. 

\begin{figure*}[!ht]
\centering
\includegraphics[width=1.0\textwidth]{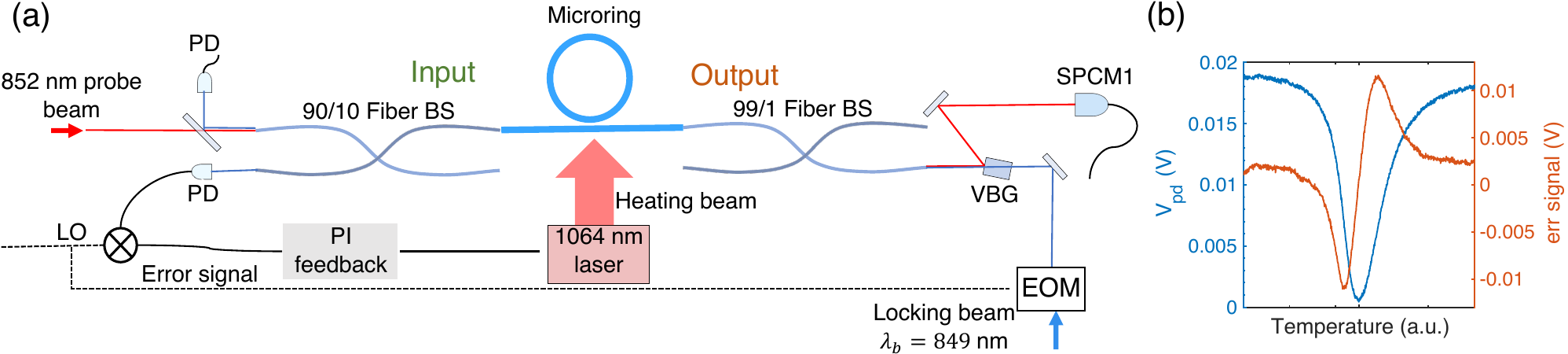}
\caption{\label{fig:fig5} (a), PDH setup for locking the resonance of the microring.
(b), Spectrum and error signal for PDH lock by adjusting the optical power of the heating beam. }
\end{figure*}

In this work, we use a `blue-color' transverse-magnetic whispering-gallery mode (WGM) for both atom trapping and microring stabilization. Here, the WGM at $\lambda_{\mathrm{b}}\approx849.552\,\mathrm{nm}$ is one free-spectral-range above the D2 line of cesium at $\lambda_\mathrm{D2}\approx852.356~$nm. We stabilize the WGM at $\lambda_{\mathrm{b}}$ using the Pound–Drever–Hall (PDH) locking technique and the probe WGM can be simultaneously stabilized at $\lambda_{\mathrm{D2}}$. As shown in Fig.~\ref{fig:fig5}, a stable locking beam with its wavelength tuned to $\lambda_{\mathrm{b}}$ is first sent through an electro-optic modulator~(EOM) with a $110~\mathrm{MHz}$ modulation frequency. After transmitting through a volume Bragg grating~(VBG), the beam is then fiber-coupled from the `Output' side via a fiber beamsplitter (BS) into the microring chip to excite the blue-color WGM. The power of the locking beam in the bus waveguide is $\approx 22~\mu$W. In the microring, it is effectively enhanced by a factor of 71~\cite{2019Optica_Chang}. We monitor the transmission using a photodetector~(PD) and mix the signal with a local oscillator (LO) to recover the error signal, as shown in Fig.~\ref{fig:fig5}(b). This provides real-time feedback to the power of a heating beam to stabilize the microring resonance.

The excited blue-color WGM provides a repulsive barrier for the microtrap used in this experiment. Prior to the probe procedure, the PDH lock is placed on hold and the power of the locking beam is ramped down to $\approx 2.2~\mu$W in 0.5~ms, reducing the potential barrier to $\Delta U_f$ as defined in the main text. Following this ramp, a probe beam with its wavelength locked at around $\lambda_\mathrm{D2}$ is sent from the `Input' side into the microring circuit. The transmitted light is directed to a single photon counting module (SPCM) for photon counting after it is deflected by the VBG. We typically send two 1~ms probe pulses, spaced by 45~ms, to measure a transmission spectrum with and without the presence of trapped atoms (only the first 500$~\mu$s of SPCM counts were used in the analyses). After the probe procedure, the locking beam is ramped back to its original power to recover the PDH lock. During a $70$~ms total hold time, there is negligible drift in the microring resonance.

\section*{Appendix B: Fictitious magnetic field and trap smoothness}
The full potential responsible for the spin-motion coupling of trapped atoms is
$\hat{V}=U_{\mathrm{tot}}(\Vec{r})\hat{I} + g_{F}\mu_\mathrm{B}\hat{F}\cdot \Vec{B}_{0} + \hat{U}^{\mathrm{v}}_{\mathrm{b}}(\Vec{r})$, 
where the first term is the total scalar potential responsible for trapping, $\hat{I}$ is the identity matrix, the second term is the Zeeman shift by the bias magnetic field, $g_F=-1/4$ is the hyperfine Land$\Acute{\mathrm{e}}$ $g$-factor for the $F=3$ ground state of cesium, $\mu_\mathrm{B}$ is the Bohr magneton, $\hat{F}$ is the total angular momentum operator, and the last term is the vector potential from the blue-color evanescent field, 
\begin{equation}
\hat{U}^{\mathrm{v}}_{\mathrm{b}}(\Vec{r})=i\frac{1}{4}\alpha^{v}\Vec{E}^*(\Vec{r})\times\Vec{E}(\Vec{r}) \cdot \frac{\hat{F}}{2F}=g_{F}\mu_\mathrm{B}\hat{F}\cdot \Vec{B}_{\mathrm{fict}} \, .
\end{equation}
Here, $\Vec{E}(\Vec{r})$ is the complex mode field, $\alpha^v$ is the atomic vector polarizability and the fictitious field follows the expression 
\begin{equation}
\Vec{B}_{\mathrm{fict}}\approx \frac{\alpha^{v}}{8g_{F}\mu_\mathrm{B} F}\xi|E(\Vec{r})|^2\hat{x}\, ,
\end{equation}
where we have used $\Vec{E}^*(\Vec{r})\times\Vec{E}(\Vec{r})\approx - i \xi|E(\Vec{r})|^2 \hat{x}$ for a mode propagating along $\hat{y}$. The coefficient
\begin{equation}
\xi = \frac{\kappa^2-4\beta^2}{\kappa^2+4\beta^2}\frac{2\mathcal{E}_y\mathcal{E}_z}{|\Vec{\mathcal{E}}|^2}\approx 0.33 \,,  
\end{equation}
takes into account the degree of circular polarization ($\sim0.98$) and the effect of mode mixing due to back-scattering in the microring \cite{2019Optica_Chang}. Here, $\kappa=\kappa_\mathrm{e}+ \kappa_\mathrm{i}$ and $\beta$ are the resonator decay rate and coherent-back scattering rate, respectively; $\Vec{\mathcal{E}}=(0,i\mathcal{E}_y,\mathcal{E}_z)$ is the electric field of the WGM normalized by $\frac{1}{2}\int\epsilon_0\epsilon(\vec{r})|\Vec{\mathcal{E}}|^2d^3r=\hbar\omega_\mathrm{b}$, where $\epsilon_0$ is the vacuum permittivity, $\epsilon(\vec{r})$ is the dielectric function, $\hbar=h/2\pi$ is the reduced Planck constant, $\omega_\mathrm{b}$ is the frequency of the WGM, and the ratio $\mathcal{E}_y/\mathcal{E}_z\approx 0.83$ is nearly uniform above the microring surface. 

Despite the presence of back-scattering, the fictitious field is still smooth along the microring and only varies along the coordinates transverse to the waveguide. For details, see discussions in Ref.~\cite{2019Optica_Chang}. 
For the scalar potential contributed by the blue-color evanescent field $U_\mathrm{b}$, there is a small sinusoidal corrugation with visibility $v=\frac{4\kappa \beta}{\kappa^2 + 4\beta^2}(1-2|\mathcal{E}_y|^2/|\vec{\mathcal{E}}|^2)\sim 20\%$ in $U_\mathrm{b}(y)\approx U_\mathrm{b}(0)[1+v\sin(2k_\mathrm{b}y)]$~\cite{2019Optica_Chang}. The effect of small corrugation should be averaged out in the atomic ensemble, given its finite temperature $T_\mathrm{trap}\approx 23~\mu$K and the smallness of $U_\mathrm{b}<5~\mu$K around the trap center $z_\mathrm{c}\approx 360\sim 440~$nm. The effect may become visible with the trapped atom temperature becoming comparable to the corrugation near the trap center. For our trap analysis along $\hat{y}$, we assume $v=0$ and trapped atoms are primarily confined by the envelope of the optical guiding beam.

\section*{Appendix C: Raman coupling for dRSC}
The fictitious field creates spin-motion Raman coupling between trap states as in Fig.~\ref{fig:fig1}(c). 
We assume the atomic motion is separable along the principal axes and expand $\hat{U}^{\mathrm{v}}_{\mathrm{b}}(\Vec{r})$ using the trapped eigenstates of $U_\mathrm{tot}(\Vec{r})\hat{I}$. The matrix element
\begin{equation}
\frac{\hbar\Omega_{\nu,\nu'} }{2}=\frac{1}{2} g_F\mu_\mathrm{B}\bra{m_F - 1}\hat{F}_{-}\ket{m_F}\bra{\nu'} B_\mathrm{fict}\ket{\nu} \,,
\end{equation}
where $\Omega_{\nu,\nu'}$ is the Raman coupling rate between trap states $\nu' < \nu$ in one given axis, and we have used $\hat{F}\cdot\hat{x}=(\hat{F}_{+}+\hat{F}_{-})/2$ with $\hat{F}_{\pm}$ being the spin raising/lowering operators. Here $\bra{\nu'}B_\mathrm{fict}\ket{\nu}$ takes finite values for all trapped states in the $z$-coordinate, as $|E(z)|^2\propto e^{-k_\mathrm{ev} z}$ decays exponentially; $\bra{\nu'_x}B_\mathrm{fict}\ket{\nu_x}$ vanishes for all odd $\Delta\nu=|\nu_x-\nu'_x|$ of transitions in the $x$-coordinate, as $|E(x)|^2 \propto \cos(qx)$. Lastly, there is no spin-motion coupling in the $y$-coordinate as $|E(y)|^2$ is constant.

\begin{figure}[!h]
\centering
\includegraphics[width=0.48\textwidth]{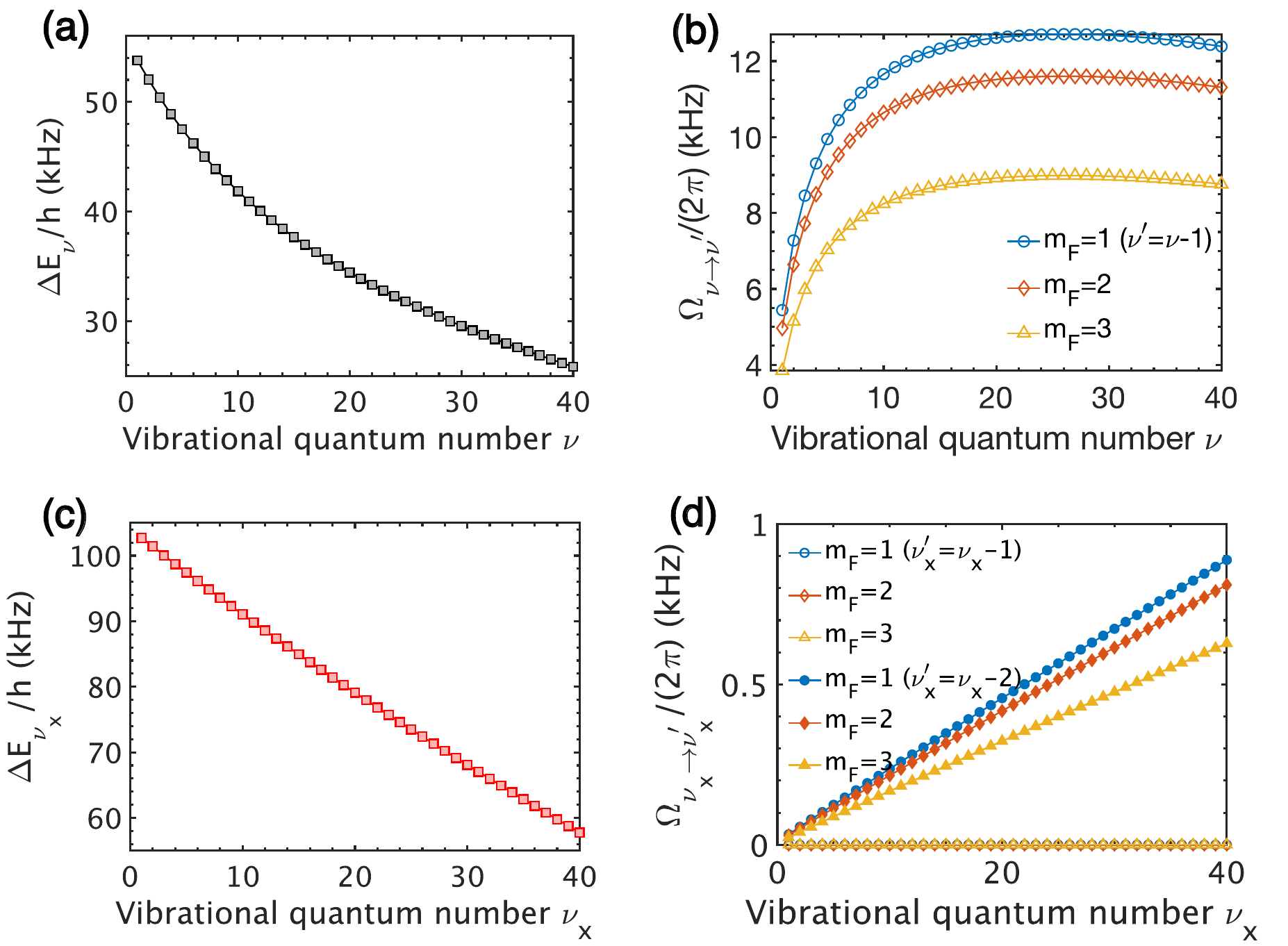}
\caption{(a), (c), Calculated level spacing for trap vibrational states along $z$ and $x$ coordinates, respectively. 
(b), (d), Calculated Raman coupling rates between the indicated states $\ket{m_{F};\nu}$ and $\ket{m_{F}-1; \nu'}$ along $z$ and $x$ coordinates, respectively. }
\label{fig:fig6}
\end{figure}
Figure~\ref{fig:fig6} shows the trap level spacing $\Delta E_{\nu}$ and sample Raman coupling rates between $\ket{m_{F};\nu}$ and $\ket{m_{F}-1; \nu-1}$ of trap states in $z$-coordinate, and rates between $\ket{m_{F};\nu_x}$ and $\ket{m_{F}+1; \nu_x-2}$ of trap states in $x$-coordinate, respectively. The Raman coupling rate between adjacent levels in $z$-axis is $\Omega/2\pi \sim 10~$kHz. Cooling is less effective along the $x$-axis, as the coupling rate is much lower $\lesssim 1~$kHz. 

\section*{Appendix D: Trap lifetime after dRSC}
Performing degenerate Raman sideband cooling (dRSC) in the $F=3$ state polarizes atoms to the lowest-energy ground state $\ket{F=3,m_F=3}$. In this state and at high in-trap densities, trapped atoms are stable against two-body inelastic collisions, but are subject to three-body recombination loss besides one-body loss caused by background gas collisions or technical heating effects. We first convert the transmission signal to the cooperativity $\bar{C}_N$ using Eq.~(\ref{eq_Transmission}). Since $C_N\propto \bar{N}$, we can fit the lifetime of the trapped atoms using $\frac{d\bar{N}}{dt} = - \bar{N}/\tau -L_3 n^2\bar{N}$, where $n$ is the in-trap density. Ignoring heating effects and given an initial density $n_0 \approx 10^{13}/$cm$^3$, we obtain a fit as shown in Fig.~2(d), giving a one-body lifetime $\tau\approx 230~$ms and a small three-body loss rate coefficient $L_3\approx 2.6\times 10^{-25}~$cm$^6$/s. Excluding the three-body term, our signal cannot be well-fitted by pure exponential decay, and the apparent lifetime is slightly shorter at $\tau^\prime\approx 150~$ms.

Once we repump atoms back to the $F=4$ state after cooling, trapped atoms are subject to fast two-body loss. It is well-known that spin-relaxation through two-body inelastic collisions could lead to trap loss~\cite{soding1998giant}. We therefore fit the transmission curve in Fig.~2(d) using $\frac{d\bar{N}}{dt} = - \bar{N}/\tau -L_2 n\bar{N}$. Assuming the same one-body lifetime $\tau\approx 230~$ms and the same initial density $n_0 \approx 10^{13}/$cm$^3$, the fitted two-body loss rate coefficient is $L_2 \approx 10^{-11}$cm$^3$/s, consistent with prior measurements using cesium atoms polarized in the $\ket{F=4,m_F=4}$ state \cite{soding1998giant}.

\section*{Appendix E: dRSC in the $F=4$ state}
\begin{figure}[!h]
\centering
\includegraphics[width=0.5\textwidth]{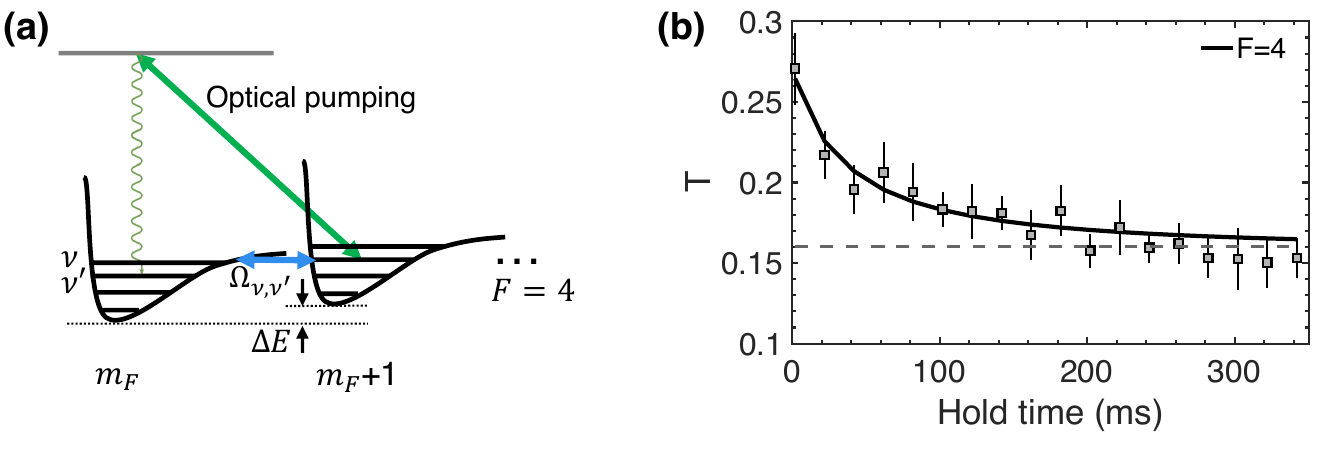}
\caption{\label{fig:fig7} dRSC for the $F=4$ state. 
(a), Similarly to cooling in the $F=3$ state, the process starts with an atom in an initial state $\ket{m_{F}; \nu}$, followed by transferring to a state $\ket{m_{F}+1; \nu^{\prime}}$. A $\sigma^{-}$ polarized optical beam can pump the atom to $\ket{m_{F}; \nu^{\prime}}$, where $\nu^{\prime}<\nu$, thus completing one cooling cycle. 
(b), Transmission curve of atoms in the $F=4$ state is shown as black squares. Black solid line is a fit. Error bars represent the standard error of the mean.
}
\end{figure}

We also performed the dRSC for atoms polarized in the $F=4$ state. Because of the positive hyperfine Land$\Acute{\mathrm{e}}$ $g$-factor for the $F=4$ state, smaller magnetic levels have lower Zeeman energy in a bias magnetic field. In this case, we perform cooling by driving the $\sigma^{-}$ transitions to optically pump atoms to the $\ket{F=4,\,m_{F}=-4}$ state, as shown in Fig.~\ref{fig:fig7}(a). This is achieved by reversing the bias field orientation from $\hat{y}$ to $-\hat{y}$ while keeping the polarization of the optical pumping beam fixed. In the experiment, we apply optical pumping using the $F=4\leftrightarrow F'=3$ transition. Following dRSC, trap loss for atoms polarized in the $\ket{F=4,m_F=-4}$ state is similarly fitted by a two-body loss model, giving $L_2 \approx  2.8\times 10^{-12}~$cm$^3$/s, consistent with a previously measured value~\cite{chin2000high}.

\section*{Appendix F: Casimir-Polder potential}
We implement finite-difference-time-domain calculations \cite{2009PRA_Casimir_rodriguez,2013NJP_1dPCW} to evaluate the Casimir-Polder potential of a ground state cesium atom above the exact dielectric structure used in the experiment, whose geometric parameters are measured from a scanning electron micrograph and an ellipsometer (for thin film thickness). The structure consists of a 950\,nm-wide and 326\,nm-thick Si$_3$N$_4$ waveguide sitting on top of a dielectric plane of 2.04\,$\mu$m-thick SiO$_2$ and 583\,nm-thick Si$_3$N$_4$ multilayer stack. In the calculation, the dielectric constant of Si$_3$N$_4$ is assumed to be 4 and that of SiO$_2$ is assumed to be $2.1025$. Both material responses are nearly constant near the transition frequencies of atomic cesium. Including material response in other frequency domains could further improve the accuracy of the Casimir-Polder potential calculation by a few percent level at $z\lesssim 100~$nm~\cite{2013NJP_1dPCW}. For longer distances, $z>  \lambda_\mathrm{D2}/2\pi$, the Casimir-Polder potential becomes less relevant because of the presence of other strong optical potentials.

The calculation is performed using a numerical grid size of 8.3~nm, and the potential is sampled in steps of 25~nm. To obtain a smooth potential function, we use an empirical fit to the numerical results 
$U_\mathrm{CP} (z) = \frac{C_{4,\mathrm{eff}}}{\tilde{z}^3 (\tilde{z}+\tilde{\lambda})}$, 
where $\tilde{\lambda} = \lambda_\mathrm{D2}/2\pi$ and $\tilde{z} = z - z_{0,\mathrm{eff}}$. Here $C_{4,\mathrm{eff}}/h = -165.36~$Hz$\cdot\mu$m$^4$ and $z_{0,\mathrm{eff}}=3.7~$nm are the only two fitting parameters. The fit residual is $|\Delta U_\mathrm{CP}|/k_\mathrm{B} \lesssim 0.6~\mu$K everywhere for $z\geq 50~$nm. Moreover, the potential remains constant over the trap range ($\Delta x \lesssim 100~$nm) along the $x$-axis due to the wide width of the waveguide.
\section*{Appendix G: Theoretical model and resonator parameters}
The transmission coefficient of single-photon transport in the atom-microring system, considering negligible coupling between a clockwise WGM and a counterclockwise WGM, has been detailed in 
Ref.~\cite{2023PRL_Coupling}. Here, we show the transmission coefficient by including a coherent back-scattering coupling in the model. When an atom is fully polarized in the stretched state ($\ket{F = 4, m_F = 4}$), it only couples to the $\sigma^{+}$ polarized probe WGM and nearly not to the counter-propagating mode ($\sigma^{-}$ polarized) due to a large asymmetry ($\sqrt{45}$ times difference) in the atom-photon coupling strength. When the probe WGM frequency $\omega_\mathrm{c}$ overlaps with the atomic resonant frequency $\omega_\mathrm{a}$, the transmission and reflection coefficients in the weak driving regime are
\begin{align}
t(\omega,g) = & 1 - \frac{2\kappa_\mathrm{e} }{\tilde{\kappa}/\tilde{\eta}} \frac{1}{1+\tilde{\eta}\tilde{C}_1} \\
r(\omega,g) = & \frac{4\kappa_\mathrm{e} }{\tilde{\kappa}/\tilde{\eta}}\frac{i\beta}{\tilde{\kappa}} \frac{1}{1+\tilde{\eta}\tilde{C}_1}\,,
\end{align}
where $\Tilde{\eta}=1/(1+\frac{\beta^2}{(\tilde{\kappa}/2)^2})$ is a reduction factor due to back-scattering, $\Tilde{\kappa}=\kappa-2i\Delta$, $\Tilde{C}_1=4g^2/(\Tilde{\kappa}\tilde{\Gamma}')$, $\tilde{\Gamma}'=\Gamma'-2i\Delta$, $\Gamma'\approx \Gamma_0$, and $\Delta=\omega-\omega_\mathrm{c}=\omega-\omega_\mathrm{a}$. 
When $N$ identical two-level atoms collectively interact through a cavity~\cite{2016PRA_GreenFunction}, they behave as a ``superatom''. The expected transmission spectrum is 
\begin{equation}\label{eq_Transmission}
T(\Delta)=|t_{N}(\omega,g)|^2=\left|1-\frac{2\kappa_{\mathrm{e}}}{\Tilde{\kappa}/\Tilde{\eta}} \frac{1}{1+\Tilde{\eta} \Tilde{C}_{N}}\right|^2\,
\end{equation}
where $\Tilde{C}_{N}=N\tilde{C}_1$ is the $N$-atom collective cooperativity; $C_N\equiv\Tilde{C}_{N}(\Delta=0)$. For the fits in Fig.~\ref{fig:fig3}(b), we use Eq.~(\ref{eq_Transmission}) and allow a small offset in $\Delta$ to account for the differential trap light shift between the ground and the excited states as well as a collective Lamb shift that will be reported in detail elsewhere. To consider shot-to-shot atom number fluctuations and variations in the coupling strength $g$ due to atoms occupying different trap states, we evaluate the averaged spectrum using a gamma distribution function of $C_N$ 
with a mean value $\bar{C}_N$. 

On the other hand, we fit the transmission spectrum for a bare microring resonator using  
$T_{0}(\Delta)=|t(\omega,g=0)|^2$ 
to extract the resonator parameters. We obtain $(\kappa_\mathrm{e}, \kappa_\mathrm{i}, \beta) = 2\pi \times (0.76, 0.94, 0.60)~$GHz. Using these results, we find the reduction factor $\eta = \tilde{\eta}(0) \approx 0.67$. We note that the back-scattering rate has increased from a negligible value reported for the same circuit in Ref.~\cite{2023PRL_Coupling}. After we optimized cooling parameters for trapping atoms and performed the experiments for an extended period of time, some amount of atoms have been adsorbed to the surface. Such an effect is under investigation.

\section*{Appendix H: Decay rate measurement}
\begin{figure}[!h]
\centering
\includegraphics[width=0.4\textwidth]{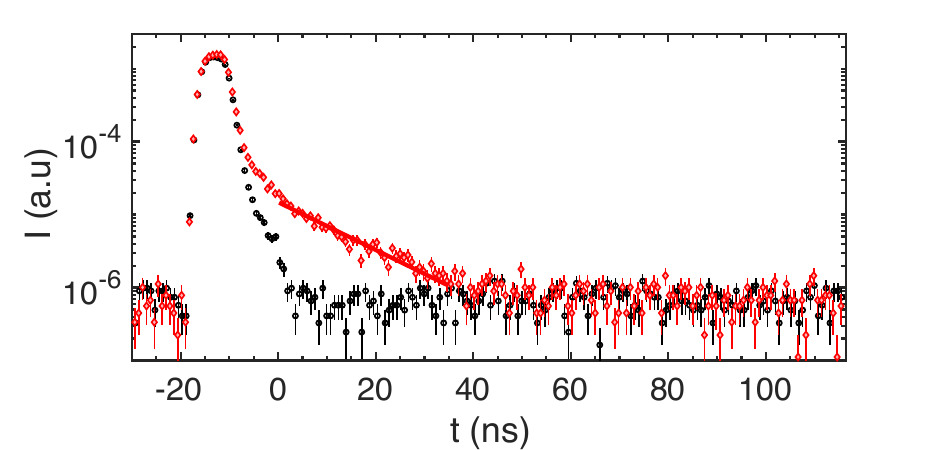}
\caption{Time-resolved transmitted photon counts without (black circle) and with (red diamonds) atoms. The red solid line shows fitted decay rate $\Gamma/\Gamma_{0}=2.33\pm0.11$ using the counts between $t=0-35\,\mathrm{ns}$ after the excitation pulse has been extinguished below the background level. Error bars represent the standard error of the mean.}
\label{fig:fig8}
\end{figure}

Pulsed excitation measurement is performed to extract the decay rate of the excited state. In the experiment, we send a weak pulse with a full width at half maximum $\sim 6\,$ns to excite the system and monitor the transmitted photon counts after the excitation pulse, as shown in Fig.~\ref{fig:fig8}. Such pulsed excitation is repeated every $200\,\mathrm{ns}$ for $2\,\mathrm{ms}$ in each experiment cycle. The total decay rate $\Gamma$ is extracted by an exponential fit to the decay curve measured between $t=0-35\,\mathrm{ns}$ after the excitation pulse has been extinguished below the background level. 
\\
\section*{Appendix I: Theoretical fits for the temperature}
In the trap-spilling procedure, we ramp down the repulsive evanescent field to force energetic atoms out of the trap. Given the short ramp and hold time, we assume no re-thermalization occurred during the procedure. The survival probability for atoms in the microtrap is evaluated using a truncated Boltzmann distribution in three spatial dimensions. Given a finite barrier height $\Delta U$, the density distribution can be evaluated~\cite{1996PRA_evaporativeCooling} as 
\begin{align}
n(\Vec{r})=n_{0}&\exp{[-(U(\Vec{r})-U_0)/k_{\mathrm{B}}T]}\times\nonumber\\
&[\mathrm{erf}(\sqrt{\kappa(\Vec{r})})-2\sqrt{\kappa(\Vec{r})/\pi}\exp{(-\kappa(\Vec{r}))}]\,
\label{eq_density_distribution}
\end{align}
where $n_0$ is the particle density at the trap minimum $U(\Vec{r}_0)=U_0$, $\kappa(\Vec{r})=[\epsilon_{t}-U(\Vec{r})]/k_{\mathrm{B}}T$, $\epsilon_{t} = \mathrm{Min}[-U_0, \Delta U]$ and $T$ is a fixed temperature. Here, we assume all trap axes have the same temperature as the saddle point in a trap-spilling potential could provide sufficient energy mixing among all spatial degrees of freedom \cite{2008PRA_Hung}. By numerically integrating Eq.~(\ref{eq_density_distribution}), we obtain the survival probability $P(\Delta U_\mathrm{min})$ for different temperatures $T$. By assuming $\bar{C}_N$ is proportional to the number of remaining atoms in the trap, we perform least $\chi^2$ fits to the data in Fig.~4 to extract the temperature of the trapped atoms. A sample fitted temperature is low at $T_\mathrm{trap} = 23(1)~\mu$K, giving mean vibrational quantum numbers $(\bar{\nu}_x, \bar{\nu}_y, \bar{\nu}_z) \approx (5,36,14)$. The root-mean-square size of atomic density distribution is $(\sigma_x,\sigma_y,\sigma_z)\approx  (94, 1916,432)~$nm along the $x,y,z$-axes, respectively. 

\begin{figure}[!h]
\centering
\includegraphics[width=0.5\textwidth]{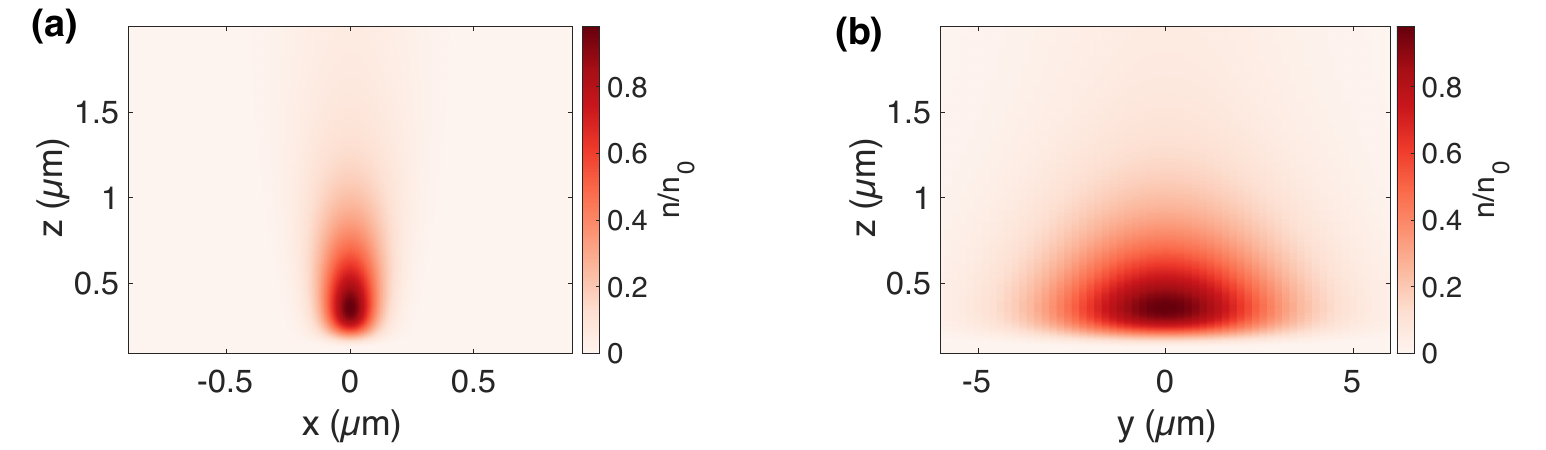}
\caption{In-trap atomic density distribution (a) $n(x,0,z)$ and (b) $n(0,y,z)$.}
\label{fig:fig9}
\end{figure}

The fit result gives 3D density distribution $n(\Vec{r})$ (see Fig.~\ref{fig:fig9}), which can be used to evaluate the single-atom cooperativity $\bar{C}_1 = 4 \bar{g}^2/\kappa\Gamma_0$, where
$\bar{g}^2 =  \frac{\int n(\Vec{r}) g^2(\Vec{r}) d^3 r}{\int n(\Vec{r}) d^3 r} \,,$
and we have obtained $\bar{g}=\sqrt{\bar{g}^2} \approx 2\pi\times 10~$MHz, leading to $\bar{C}_1 \approx 0.05$. Given our fitted temperature range, $\bar{C}_1$ varies by less than 1\%. 

\bibliography{apssamp}

\begin{thebibliography}{68}%
\makeatletter
\providecommand \@ifxundefined [1]{%
 \@ifx{#1\undefined}
}%
\providecommand \@ifnum [1]{%
 \ifnum #1\expandafter \@firstoftwo
 \else \expandafter \@secondoftwo
 \fi
}%
\providecommand \@ifx [1]{%
 \ifx #1\expandafter \@firstoftwo
 \else \expandafter \@secondoftwo
 \fi
}%
\providecommand \natexlab [1]{#1}%
\providecommand \enquote  [1]{``#1''}%
\providecommand \bibnamefont  [1]{#1}%
\providecommand \bibfnamefont [1]{#1}%
\providecommand \citenamefont [1]{#1}%
\providecommand \href@noop [0]{\@secondoftwo}%
\providecommand \href [0]{\begingroup \@sanitize@url \@href}%
\providecommand \@href[1]{\@@startlink{#1}\@@href}%
\providecommand \@@href[1]{\endgroup#1\@@endlink}%
\providecommand \@sanitize@url [0]{\catcode `\\12\catcode `\$12\catcode `\&12\catcode `\#12\catcode `\^12\catcode `\_12\catcode `\%12\relax}%
\providecommand \@@startlink[1]{}%
\providecommand \@@endlink[0]{}%
\providecommand \url  [0]{\begingroup\@sanitize@url \@url }%
\providecommand \@url [1]{\endgroup\@href {#1}{\urlprefix }}%
\providecommand \urlprefix  [0]{URL }%
\providecommand \Eprint [0]{\href }%
\providecommand \doibase [0]{https://doi.org/}%
\providecommand \selectlanguage [0]{\@gobble}%
\providecommand \bibinfo  [0]{\@secondoftwo}%
\providecommand \bibfield  [0]{\@secondoftwo}%
\providecommand \translation [1]{[#1]}%
\providecommand \BibitemOpen [0]{}%
\providecommand \bibitemStop [0]{}%
\providecommand \bibitemNoStop [0]{.\EOS\space}%
\providecommand \EOS [0]{\spacefactor3000\relax}%
\providecommand \BibitemShut  [1]{\csname bibitem#1\endcsname}%
\let\auto@bib@innerbib\@empty
\bibitem [{\citenamefont {Chang}\ \emph {et~al.}(2018)\citenamefont {Chang}, \citenamefont {Douglas}, \citenamefont {Gonz\'alez-Tudela}, \citenamefont {Hung},\ and\ \citenamefont {Kimble}}]{2018RMP_DEChang}%
  \BibitemOpen
  \bibfield  {author} {\bibinfo {author} {\bibfnamefont {D.~E.}\ \bibnamefont {Chang}}, \bibinfo {author} {\bibfnamefont {J.~S.}\ \bibnamefont {Douglas}}, \bibinfo {author} {\bibfnamefont {A.}~\bibnamefont {Gonz\'alez-Tudela}}, \bibinfo {author} {\bibfnamefont {C.-L.}\ \bibnamefont {Hung}},\ and\ \bibinfo {author} {\bibfnamefont {H.~J.}\ \bibnamefont {Kimble}},\ }\bibfield  {title} {\bibinfo {title} {Colloquium: Quantum matter built from nanoscopic lattices of atoms and photons},\ }\href {https://doi.org/10.1103/RevModPhys.90.031002} {\bibfield  {journal} {\bibinfo  {journal} {Rev. Mod. Phys.}\ }\textbf {\bibinfo {volume} {90}},\ \bibinfo {pages} {031002} (\bibinfo {year} {2018})}\BibitemShut {NoStop}%
\bibitem [{\citenamefont {Chang}\ \emph {et~al.}(2014)\citenamefont {Chang}, \citenamefont {Vuleti{\'{c}}},\ and\ \citenamefont {Lukin}}]{2014NaturePhotonics_chang}%
  \BibitemOpen
  \bibfield  {author} {\bibinfo {author} {\bibfnamefont {D.~E.}\ \bibnamefont {Chang}}, \bibinfo {author} {\bibfnamefont {V.}~\bibnamefont {Vuleti{\'{c}}}},\ and\ \bibinfo {author} {\bibfnamefont {M.~D.}\ \bibnamefont {Lukin}},\ }\bibfield  {title} {\bibinfo {title} {Quantum nonlinear optics---photon by photon},\ }\href {https://doi.org/10.1038/nphoton.2014.192} {\bibfield  {journal} {\bibinfo  {journal} {Nature Photonics}\ }\textbf {\bibinfo {volume} {8}},\ \bibinfo {pages} {685} (\bibinfo {year} {2014})}\BibitemShut {NoStop}%
\bibitem [{\citenamefont {Kimble}(2008)}]{2008Nature_KimbleInternet}%
  \BibitemOpen
  \bibfield  {author} {\bibinfo {author} {\bibfnamefont {H.~J.}\ \bibnamefont {Kimble}},\ }\bibfield  {title} {\bibinfo {title} {The quantum internet},\ }\href {https://doi.org/10.1038/nature07127} {\bibfield  {journal} {\bibinfo  {journal} {Nature}\ }\textbf {\bibinfo {volume} {453}},\ \bibinfo {pages} {1023} (\bibinfo {year} {2008})}\BibitemShut {NoStop}%
\bibitem [{\citenamefont {Paulisch}\ \emph {et~al.}(2019)\citenamefont {Paulisch}, \citenamefont {Perarnau-Llobet}, \citenamefont {Gonz\'alez-Tudela},\ and\ \citenamefont {Cirac}}]{2019PRA_metrology}%
  \BibitemOpen
  \bibfield  {author} {\bibinfo {author} {\bibfnamefont {V.}~\bibnamefont {Paulisch}}, \bibinfo {author} {\bibfnamefont {M.}~\bibnamefont {Perarnau-Llobet}}, \bibinfo {author} {\bibfnamefont {A.}~\bibnamefont {Gonz\'alez-Tudela}},\ and\ \bibinfo {author} {\bibfnamefont {J.~I.}\ \bibnamefont {Cirac}},\ }\bibfield  {title} {\bibinfo {title} {Quantum metrology with one-dimensional superradiant photonic states},\ }\href {https://doi.org/10.1103/PhysRevA.99.043807} {\bibfield  {journal} {\bibinfo  {journal} {Phys. Rev. A}\ }\textbf {\bibinfo {volume} {99}},\ \bibinfo {pages} {043807} (\bibinfo {year} {2019})}\BibitemShut {NoStop}%
\bibitem [{\citenamefont {Douglas}\ \emph {et~al.}(2016)\citenamefont {Douglas}, \citenamefont {Caneva},\ and\ \citenamefont {Chang}}]{2016PRX_PhotonMolecules}%
  \BibitemOpen
  \bibfield  {author} {\bibinfo {author} {\bibfnamefont {J.~S.}\ \bibnamefont {Douglas}}, \bibinfo {author} {\bibfnamefont {T.}~\bibnamefont {Caneva}},\ and\ \bibinfo {author} {\bibfnamefont {D.~E.}\ \bibnamefont {Chang}},\ }\bibfield  {title} {\bibinfo {title} {Photon molecules in atomic gases trapped near photonic crystal waveguides},\ }\href {https://doi.org/10.1103/PhysRevX.6.031017} {\bibfield  {journal} {\bibinfo  {journal} {Phys. Rev. X}\ }\textbf {\bibinfo {volume} {6}},\ \bibinfo {pages} {031017} (\bibinfo {year} {2016})}\BibitemShut {NoStop}%
\bibitem [{\citenamefont {Shomroni}\ \emph {et~al.}(2014)\citenamefont {Shomroni}, \citenamefont {Rosenblum}, \citenamefont {Lovsky}, \citenamefont {Bechler}, \citenamefont {Guendelman},\ and\ \citenamefont {Dayan}}]{2014Science_Dayan}%
  \BibitemOpen
  \bibfield  {author} {\bibinfo {author} {\bibfnamefont {I.}~\bibnamefont {Shomroni}}, \bibinfo {author} {\bibfnamefont {S.}~\bibnamefont {Rosenblum}}, \bibinfo {author} {\bibfnamefont {Y.}~\bibnamefont {Lovsky}}, \bibinfo {author} {\bibfnamefont {O.}~\bibnamefont {Bechler}}, \bibinfo {author} {\bibfnamefont {G.}~\bibnamefont {Guendelman}},\ and\ \bibinfo {author} {\bibfnamefont {B.}~\bibnamefont {Dayan}},\ }\bibfield  {title} {\bibinfo {title} {All-optical routing of single photons by a one-atom switch controlled by a single photon},\ }\href {https://doi.org/10.1126/science.1254699} {\bibfield  {journal} {\bibinfo  {journal} {Science}\ }\textbf {\bibinfo {volume} {345}},\ \bibinfo {pages} {903} (\bibinfo {year} {2014})}\BibitemShut {NoStop}%
\bibitem [{\citenamefont {Le~Jeannic}\ \emph {et~al.}(2022)\citenamefont {Le~Jeannic}, \citenamefont {Tiranov}, \citenamefont {Carolan}, \citenamefont {Ramos}, \citenamefont {Wang}, \citenamefont {Appel}, \citenamefont {Scholz}, \citenamefont {Wieck}, \citenamefont {Ludwig}, \citenamefont {Rotenberg}, \citenamefont {Midolo}, \citenamefont {Garc{\'i}a-Ripoll}, \citenamefont {S{\o}rensen},\ and\ \citenamefont {Lodahl}}]{2022NaturePhysics_QD}%
  \BibitemOpen
  \bibfield  {author} {\bibinfo {author} {\bibfnamefont {H.}~\bibnamefont {Le~Jeannic}}, \bibinfo {author} {\bibfnamefont {A.}~\bibnamefont {Tiranov}}, \bibinfo {author} {\bibfnamefont {J.}~\bibnamefont {Carolan}}, \bibinfo {author} {\bibfnamefont {T.}~\bibnamefont {Ramos}}, \bibinfo {author} {\bibfnamefont {Y.}~\bibnamefont {Wang}}, \bibinfo {author} {\bibfnamefont {M.~H.}\ \bibnamefont {Appel}}, \bibinfo {author} {\bibfnamefont {S.}~\bibnamefont {Scholz}}, \bibinfo {author} {\bibfnamefont {A.~D.}\ \bibnamefont {Wieck}}, \bibinfo {author} {\bibfnamefont {A.}~\bibnamefont {Ludwig}}, \bibinfo {author} {\bibfnamefont {N.}~\bibnamefont {Rotenberg}}, \bibinfo {author} {\bibfnamefont {L.}~\bibnamefont {Midolo}}, \bibinfo {author} {\bibfnamefont {J.~J.}\ \bibnamefont {Garc{\'i}a-Ripoll}}, \bibinfo {author} {\bibfnamefont {A.~S.}\ \bibnamefont {S{\o}rensen}},\ and\ \bibinfo {author} {\bibfnamefont {P.}~\bibnamefont {Lodahl}},\ }\bibfield  {title} {\bibinfo {title} {Dynamical photon--photon interaction mediated by a
  quantum emitter},\ }\href {https://doi.org/10.1038/s41567-022-01720-x} {\bibfield  {journal} {\bibinfo  {journal} {Nature Physics}\ }\textbf {\bibinfo {volume} {18}},\ \bibinfo {pages} {1191} (\bibinfo {year} {2022})}\BibitemShut {NoStop}%
\bibitem [{\citenamefont {Tomm}\ \emph {et~al.}(2023)\citenamefont {Tomm}, \citenamefont {Mahmoodian}, \citenamefont {Antoniadis}, \citenamefont {Schott}, \citenamefont {Valentin}, \citenamefont {Wieck}, \citenamefont {Ludwig}, \citenamefont {Javadi},\ and\ \citenamefont {Warburton}}]{2023NaturePhysics_photonboundstate}%
  \BibitemOpen
  \bibfield  {author} {\bibinfo {author} {\bibfnamefont {N.}~\bibnamefont {Tomm}}, \bibinfo {author} {\bibfnamefont {S.}~\bibnamefont {Mahmoodian}}, \bibinfo {author} {\bibfnamefont {N.~O.}\ \bibnamefont {Antoniadis}}, \bibinfo {author} {\bibfnamefont {R.}~\bibnamefont {Schott}}, \bibinfo {author} {\bibfnamefont {S.~R.}\ \bibnamefont {Valentin}}, \bibinfo {author} {\bibfnamefont {A.~D.}\ \bibnamefont {Wieck}}, \bibinfo {author} {\bibfnamefont {A.}~\bibnamefont {Ludwig}}, \bibinfo {author} {\bibfnamefont {A.}~\bibnamefont {Javadi}},\ and\ \bibinfo {author} {\bibfnamefont {R.~J.}\ \bibnamefont {Warburton}},\ }\bibfield  {title} {\bibinfo {title} {Photon bound state dynamics from a single artificial atom},\ }\href {https://doi.org/10.1038/s41567-023-01997-6} {\bibfield  {journal} {\bibinfo  {journal} {Nature Physics}\ }\textbf {\bibinfo {volume} {19}},\ \bibinfo {pages} {857} (\bibinfo {year} {2023})}\BibitemShut {NoStop}%
\bibitem [{\citenamefont {Sheremet}\ \emph {et~al.}(2023)\citenamefont {Sheremet}, \citenamefont {Petrov}, \citenamefont {Iorsh}, \citenamefont {Poshakinskiy},\ and\ \citenamefont {Poddubny}}]{2023RMP_wQED}%
  \BibitemOpen
  \bibfield  {author} {\bibinfo {author} {\bibfnamefont {A.~S.}\ \bibnamefont {Sheremet}}, \bibinfo {author} {\bibfnamefont {M.~I.}\ \bibnamefont {Petrov}}, \bibinfo {author} {\bibfnamefont {I.~V.}\ \bibnamefont {Iorsh}}, \bibinfo {author} {\bibfnamefont {A.~V.}\ \bibnamefont {Poshakinskiy}},\ and\ \bibinfo {author} {\bibfnamefont {A.~N.}\ \bibnamefont {Poddubny}},\ }\bibfield  {title} {\bibinfo {title} {Waveguide quantum electrodynamics: Collective radiance and photon-photon correlations},\ }\href {https://doi.org/10.1103/RevModPhys.95.015002} {\bibfield  {journal} {\bibinfo  {journal} {Rev. Mod. Phys.}\ }\textbf {\bibinfo {volume} {95}},\ \bibinfo {pages} {015002} (\bibinfo {year} {2023})}\BibitemShut {NoStop}%
\bibitem [{\citenamefont {Goban}\ \emph {et~al.}(2015)\citenamefont {Goban}, \citenamefont {Hung}, \citenamefont {Hood}, \citenamefont {Yu}, \citenamefont {Muniz}, \citenamefont {Painter},\ and\ \citenamefont {Kimble}}]{2015PRL_pcw}%
  \BibitemOpen
  \bibfield  {author} {\bibinfo {author} {\bibfnamefont {A.}~\bibnamefont {Goban}}, \bibinfo {author} {\bibfnamefont {C.-L.}\ \bibnamefont {Hung}}, \bibinfo {author} {\bibfnamefont {J.~D.}\ \bibnamefont {Hood}}, \bibinfo {author} {\bibfnamefont {S.-P.}\ \bibnamefont {Yu}}, \bibinfo {author} {\bibfnamefont {J.~A.}\ \bibnamefont {Muniz}}, \bibinfo {author} {\bibfnamefont {O.}~\bibnamefont {Painter}},\ and\ \bibinfo {author} {\bibfnamefont {H.~J.}\ \bibnamefont {Kimble}},\ }\bibfield  {title} {\bibinfo {title} {Superradiance for atoms trapped along a photonic crystal waveguide},\ }\href {https://doi.org/10.1103/PhysRevLett.115.063601} {\bibfield  {journal} {\bibinfo  {journal} {Phys. Rev. Lett.}\ }\textbf {\bibinfo {volume} {115}},\ \bibinfo {pages} {063601} (\bibinfo {year} {2015})}\BibitemShut {NoStop}%
\bibitem [{\citenamefont {Solano}\ \emph {et~al.}(2017)\citenamefont {Solano}, \citenamefont {Barberis-Blostein}, \citenamefont {Fatemi}, \citenamefont {Orozco},\ and\ \citenamefont {Rolston}}]{2017NC_superradiance}%
  \BibitemOpen
  \bibfield  {author} {\bibinfo {author} {\bibfnamefont {P.}~\bibnamefont {Solano}}, \bibinfo {author} {\bibfnamefont {P.}~\bibnamefont {Barberis-Blostein}}, \bibinfo {author} {\bibfnamefont {F.~K.}\ \bibnamefont {Fatemi}}, \bibinfo {author} {\bibfnamefont {L.~A.}\ \bibnamefont {Orozco}},\ and\ \bibinfo {author} {\bibfnamefont {S.~L.}\ \bibnamefont {Rolston}},\ }\bibfield  {title} {\bibinfo {title} {Super-radiance reveals infinite-range dipole interactions through a nanofiber},\ }\href {https://doi.org/10.1038/s41467-017-01994-3} {\bibfield  {journal} {\bibinfo  {journal} {Nature Communications}\ }\textbf {\bibinfo {volume} {8}},\ \bibinfo {pages} {1857} (\bibinfo {year} {2017})}\BibitemShut {NoStop}%
\bibitem [{\citenamefont {Pennetta}\ \emph {et~al.}(2022)\citenamefont {Pennetta}, \citenamefont {Blaha}, \citenamefont {Johnson}, \citenamefont {Lechner}, \citenamefont {Schneeweiss}, \citenamefont {Volz},\ and\ \citenamefont {Rauschenbeutel}}]{2022PRL_collecticedecay}%
  \BibitemOpen
  \bibfield  {author} {\bibinfo {author} {\bibfnamefont {R.}~\bibnamefont {Pennetta}}, \bibinfo {author} {\bibfnamefont {M.}~\bibnamefont {Blaha}}, \bibinfo {author} {\bibfnamefont {A.}~\bibnamefont {Johnson}}, \bibinfo {author} {\bibfnamefont {D.}~\bibnamefont {Lechner}}, \bibinfo {author} {\bibfnamefont {P.}~\bibnamefont {Schneeweiss}}, \bibinfo {author} {\bibfnamefont {J.}~\bibnamefont {Volz}},\ and\ \bibinfo {author} {\bibfnamefont {A.}~\bibnamefont {Rauschenbeutel}},\ }\bibfield  {title} {\bibinfo {title} {Collective radiative dynamics of an ensemble of cold atoms coupled to an optical waveguide},\ }\href {https://doi.org/10.1103/PhysRevLett.128.073601} {\bibfield  {journal} {\bibinfo  {journal} {Phys. Rev. Lett.}\ }\textbf {\bibinfo {volume} {128}},\ \bibinfo {pages} {073601} (\bibinfo {year} {2022})}\BibitemShut {NoStop}%
\bibitem [{\citenamefont {Cardenas-Lopez}\ \emph {et~al.}(2023)\citenamefont {Cardenas-Lopez}, \citenamefont {Masson}, \citenamefont {Zager},\ and\ \citenamefont {Asenjo-Garcia}}]{2023PRL_SuperradiancewQED}%
  \BibitemOpen
  \bibfield  {author} {\bibinfo {author} {\bibfnamefont {S.}~\bibnamefont {Cardenas-Lopez}}, \bibinfo {author} {\bibfnamefont {S.~J.}\ \bibnamefont {Masson}}, \bibinfo {author} {\bibfnamefont {Z.}~\bibnamefont {Zager}},\ and\ \bibinfo {author} {\bibfnamefont {A.}~\bibnamefont {Asenjo-Garcia}},\ }\bibfield  {title} {\bibinfo {title} {Many-body superradiance and dynamical mirror symmetry breaking in waveguide {QED}},\ }\href {https://doi.org/10.1103/PhysRevLett.131.033605} {\bibfield  {journal} {\bibinfo  {journal} {Phys. Rev. Lett.}\ }\textbf {\bibinfo {volume} {131}},\ \bibinfo {pages} {033605} (\bibinfo {year} {2023})}\BibitemShut {NoStop}%
\bibitem [{\citenamefont {Sinha}\ \emph {et~al.}(2020)\citenamefont {Sinha}, \citenamefont {Meystre}, \citenamefont {Goldschmidt}, \citenamefont {Fatemi}, \citenamefont {Rolston},\ and\ \citenamefont {Solano}}]{2020PRL_Kanu}%
  \BibitemOpen
  \bibfield  {author} {\bibinfo {author} {\bibfnamefont {K.}~\bibnamefont {Sinha}}, \bibinfo {author} {\bibfnamefont {P.}~\bibnamefont {Meystre}}, \bibinfo {author} {\bibfnamefont {E.~A.}\ \bibnamefont {Goldschmidt}}, \bibinfo {author} {\bibfnamefont {F.~K.}\ \bibnamefont {Fatemi}}, \bibinfo {author} {\bibfnamefont {S.~L.}\ \bibnamefont {Rolston}},\ and\ \bibinfo {author} {\bibfnamefont {P.}~\bibnamefont {Solano}},\ }\bibfield  {title} {\bibinfo {title} {Non-{M}arkovian collective emission from macroscopically separated emitters},\ }\href {https://doi.org/10.1103/PhysRevLett.124.043603} {\bibfield  {journal} {\bibinfo  {journal} {Phys. Rev. Lett.}\ }\textbf {\bibinfo {volume} {124}},\ \bibinfo {pages} {043603} (\bibinfo {year} {2020})}\BibitemShut {NoStop}%
\bibitem [{\citenamefont {Zhang}(2023)}]{2023PRL_ZenoRegime}%
  \BibitemOpen
  \bibfield  {author} {\bibinfo {author} {\bibfnamefont {Y.-X.}\ \bibnamefont {Zhang}},\ }\bibfield  {title} {\bibinfo {title} {Zeno regime of collective emission: Non-{M}arkovianity beyond retardation},\ }\href {https://doi.org/10.1103/PhysRevLett.131.193603} {\bibfield  {journal} {\bibinfo  {journal} {Phys. Rev. Lett.}\ }\textbf {\bibinfo {volume} {131}},\ \bibinfo {pages} {193603} (\bibinfo {year} {2023})}\BibitemShut {NoStop}%
\bibitem [{\citenamefont {Douglas}\ \emph {et~al.}(2015)\citenamefont {Douglas}, \citenamefont {Habibian}, \citenamefont {Hung}, \citenamefont {Gorshkov}, \citenamefont {Kimble},\ and\ \citenamefont {Chang}}]{2015NaturePhotonics_manybodymodels}%
  \BibitemOpen
  \bibfield  {author} {\bibinfo {author} {\bibfnamefont {J.~S.}\ \bibnamefont {Douglas}}, \bibinfo {author} {\bibfnamefont {H.}~\bibnamefont {Habibian}}, \bibinfo {author} {\bibfnamefont {C.-L.}\ \bibnamefont {Hung}}, \bibinfo {author} {\bibfnamefont {A.~V.}\ \bibnamefont {Gorshkov}}, \bibinfo {author} {\bibfnamefont {H.~J.}\ \bibnamefont {Kimble}},\ and\ \bibinfo {author} {\bibfnamefont {D.~E.}\ \bibnamefont {Chang}},\ }\bibfield  {title} {\bibinfo {title} {Quantum many-body models with cold atoms coupled to photonic crystals},\ }\href {https://doi.org/10.1038/nphoton.2015.57} {\bibfield  {journal} {\bibinfo  {journal} {Nature Photonics}\ }\textbf {\bibinfo {volume} {9}},\ \bibinfo {pages} {326} (\bibinfo {year} {2015})}\BibitemShut {NoStop}%
\bibitem [{\citenamefont {Gonz{\'a}lez-Tudela}\ \emph {et~al.}(2015)\citenamefont {Gonz{\'a}lez-Tudela}, \citenamefont {Hung}, \citenamefont {Chang}, \citenamefont {Cirac},\ and\ \citenamefont {Kimble}}]{2015NaturePhotonics_subwavelength}%
  \BibitemOpen
  \bibfield  {author} {\bibinfo {author} {\bibfnamefont {A.}~\bibnamefont {Gonz{\'a}lez-Tudela}}, \bibinfo {author} {\bibfnamefont {C.-L.}\ \bibnamefont {Hung}}, \bibinfo {author} {\bibfnamefont {D.~E.}\ \bibnamefont {Chang}}, \bibinfo {author} {\bibfnamefont {J.~I.}\ \bibnamefont {Cirac}},\ and\ \bibinfo {author} {\bibfnamefont {H.~J.}\ \bibnamefont {Kimble}},\ }\bibfield  {title} {\bibinfo {title} {Subwavelength vacuum lattices and atom--atom interactions in two-dimensional photonic crystals},\ }\href {https://doi.org/10.1038/nphoton.2015.54} {\bibfield  {journal} {\bibinfo  {journal} {Nature Photonics}\ }\textbf {\bibinfo {volume} {9}},\ \bibinfo {pages} {320} (\bibinfo {year} {2015})}\BibitemShut {NoStop}%
\bibitem [{\citenamefont {Hung}\ \emph {et~al.}(2016)\citenamefont {Hung}, \citenamefont {González-Tudela}, \citenamefont {Cirac},\ and\ \citenamefont {Kimble}}]{2016PNAS_hung}%
  \BibitemOpen
  \bibfield  {author} {\bibinfo {author} {\bibfnamefont {C.-L.}\ \bibnamefont {Hung}}, \bibinfo {author} {\bibfnamefont {A.}~\bibnamefont {González-Tudela}}, \bibinfo {author} {\bibfnamefont {J.~I.}\ \bibnamefont {Cirac}},\ and\ \bibinfo {author} {\bibfnamefont {H.~J.}\ \bibnamefont {Kimble}},\ }\bibfield  {title} {\bibinfo {title} {Quantum spin dynamics with pairwise-tunable, long-range interactions},\ }\href {https://doi.org/10.1073/pnas.1603777113} {\bibfield  {journal} {\bibinfo  {journal} {Proceedings of the National Academy of Sciences}\ }\textbf {\bibinfo {volume} {113}},\ \bibinfo {pages} {E4946} (\bibinfo {year} {2016})}\BibitemShut {NoStop}%
\bibitem [{\citenamefont {Hood}\ \emph {et~al.}(2016)\citenamefont {Hood}, \citenamefont {Goban}, \citenamefont {Asenjo-Garcia}, \citenamefont {Lu}, \citenamefont {Yu}, \citenamefont {Chang},\ and\ \citenamefont {Kimble}}]{2016PNAS_Hood_badedge}%
  \BibitemOpen
  \bibfield  {author} {\bibinfo {author} {\bibfnamefont {J.~D.}\ \bibnamefont {Hood}}, \bibinfo {author} {\bibfnamefont {A.}~\bibnamefont {Goban}}, \bibinfo {author} {\bibfnamefont {A.}~\bibnamefont {Asenjo-Garcia}}, \bibinfo {author} {\bibfnamefont {M.}~\bibnamefont {Lu}}, \bibinfo {author} {\bibfnamefont {S.-P.}\ \bibnamefont {Yu}}, \bibinfo {author} {\bibfnamefont {D.~E.}\ \bibnamefont {Chang}},\ and\ \bibinfo {author} {\bibfnamefont {H.~J.}\ \bibnamefont {Kimble}},\ }\bibfield  {title} {\bibinfo {title} {Atom–atom interactions around the band edge of a photonic crystal waveguide},\ }\href {https://doi.org/10.1073/pnas.1603788113} {\bibfield  {journal} {\bibinfo  {journal} {Proceedings of the National Academy of Sciences}\ }\textbf {\bibinfo {volume} {113}},\ \bibinfo {pages} {10507} (\bibinfo {year} {2016})}\BibitemShut {NoStop}%
\bibitem [{\citenamefont {Bello}\ \emph {et~al.}(2019)\citenamefont {Bello}, \citenamefont {Platero}, \citenamefont {Cirac},\ and\ \citenamefont {González-Tudela}}]{2019ScienceAdvances_topologicalwQED}%
  \BibitemOpen
  \bibfield  {author} {\bibinfo {author} {\bibfnamefont {M.}~\bibnamefont {Bello}}, \bibinfo {author} {\bibfnamefont {G.}~\bibnamefont {Platero}}, \bibinfo {author} {\bibfnamefont {J.~I.}\ \bibnamefont {Cirac}},\ and\ \bibinfo {author} {\bibfnamefont {A.}~\bibnamefont {González-Tudela}},\ }\bibfield  {title} {\bibinfo {title} {Unconventional quantum optics in topological waveguide {QED}},\ }\href {https://doi.org/10.1126/sciadv.aaw0297} {\bibfield  {journal} {\bibinfo  {journal} {Science Advances}\ }\textbf {\bibinfo {volume} {5}},\ \bibinfo {pages} {eaaw0297} (\bibinfo {year} {2019})}\BibitemShut {NoStop}%
\bibitem [{\citenamefont {Periwal}\ \emph {et~al.}(2021)\citenamefont {Periwal}, \citenamefont {Cooper}, \citenamefont {Kunkel}, \citenamefont {Wienand}, \citenamefont {Davis},\ and\ \citenamefont {Schleier-Smith}}]{2021Nature_ProgramableInteraction_monika}%
  \BibitemOpen
  \bibfield  {author} {\bibinfo {author} {\bibfnamefont {A.}~\bibnamefont {Periwal}}, \bibinfo {author} {\bibfnamefont {E.~S.}\ \bibnamefont {Cooper}}, \bibinfo {author} {\bibfnamefont {P.}~\bibnamefont {Kunkel}}, \bibinfo {author} {\bibfnamefont {J.~F.}\ \bibnamefont {Wienand}}, \bibinfo {author} {\bibfnamefont {E.~J.}\ \bibnamefont {Davis}},\ and\ \bibinfo {author} {\bibfnamefont {M.}~\bibnamefont {Schleier-Smith}},\ }\bibfield  {title} {\bibinfo {title} {Programmable interactions and emergent geometry in an array of atom clouds},\ }\href {https://doi.org/10.1038/s41586-021-04156-0} {\bibfield  {journal} {\bibinfo  {journal} {Nature}\ }\textbf {\bibinfo {volume} {600}},\ \bibinfo {pages} {630} (\bibinfo {year} {2021})}\BibitemShut {NoStop}%
\bibitem [{\citenamefont {Ramos}\ \emph {et~al.}(2014)\citenamefont {Ramos}, \citenamefont {Pichler}, \citenamefont {Daley},\ and\ \citenamefont {Zoller}}]{2014PRL_QuantumSpinDimers}%
  \BibitemOpen
  \bibfield  {author} {\bibinfo {author} {\bibfnamefont {T.}~\bibnamefont {Ramos}}, \bibinfo {author} {\bibfnamefont {H.}~\bibnamefont {Pichler}}, \bibinfo {author} {\bibfnamefont {A.~J.}\ \bibnamefont {Daley}},\ and\ \bibinfo {author} {\bibfnamefont {P.}~\bibnamefont {Zoller}},\ }\bibfield  {title} {\bibinfo {title} {Quantum spin dimers from chiral dissipation in cold-atom chains},\ }\href {https://doi.org/10.1103/PhysRevLett.113.237203} {\bibfield  {journal} {\bibinfo  {journal} {Phys. Rev. Lett.}\ }\textbf {\bibinfo {volume} {113}},\ \bibinfo {pages} {237203} (\bibinfo {year} {2014})}\BibitemShut {NoStop}%
\bibitem [{\citenamefont {Asenjo-Garcia}\ \emph {et~al.}(2017{\natexlab{a}})\citenamefont {Asenjo-Garcia}, \citenamefont {Moreno-Cardoner}, \citenamefont {Albrecht}, \citenamefont {Kimble},\ and\ \citenamefont {Chang}}]{2017PRX_ana_photonstorage}%
  \BibitemOpen
  \bibfield  {author} {\bibinfo {author} {\bibfnamefont {A.}~\bibnamefont {Asenjo-Garcia}}, \bibinfo {author} {\bibfnamefont {M.}~\bibnamefont {Moreno-Cardoner}}, \bibinfo {author} {\bibfnamefont {A.}~\bibnamefont {Albrecht}}, \bibinfo {author} {\bibfnamefont {H.~J.}\ \bibnamefont {Kimble}},\ and\ \bibinfo {author} {\bibfnamefont {D.~E.}\ \bibnamefont {Chang}},\ }\bibfield  {title} {\bibinfo {title} {Exponential improvement in photon storage fidelities using subradiance and ``selective radiance'' in atomic arrays},\ }\href {https://doi.org/10.1103/PhysRevX.7.031024} {\bibfield  {journal} {\bibinfo  {journal} {Phys. Rev. X}\ }\textbf {\bibinfo {volume} {7}},\ \bibinfo {pages} {031024} (\bibinfo {year} {2017}{\natexlab{a}})}\BibitemShut {NoStop}%
\bibitem [{\citenamefont {Gonz\'alez-Tudela}\ \emph {et~al.}(2017)\citenamefont {Gonz\'alez-Tudela}, \citenamefont {Paulisch}, \citenamefont {Kimble},\ and\ \citenamefont {Cirac}}]{2017PRL_multiphotongeneration}%
  \BibitemOpen
  \bibfield  {author} {\bibinfo {author} {\bibfnamefont {A.}~\bibnamefont {Gonz\'alez-Tudela}}, \bibinfo {author} {\bibfnamefont {V.}~\bibnamefont {Paulisch}}, \bibinfo {author} {\bibfnamefont {H.~J.}\ \bibnamefont {Kimble}},\ and\ \bibinfo {author} {\bibfnamefont {J.~I.}\ \bibnamefont {Cirac}},\ }\bibfield  {title} {\bibinfo {title} {Efficient multiphoton generation in waveguide quantum electrodynamics},\ }\href {https://doi.org/10.1103/PhysRevLett.118.213601} {\bibfield  {journal} {\bibinfo  {journal} {Phys. Rev. Lett.}\ }\textbf {\bibinfo {volume} {118}},\ \bibinfo {pages} {213601} (\bibinfo {year} {2017})}\BibitemShut {NoStop}%
\bibitem [{\citenamefont {Haas}\ \emph {et~al.}(2014)\citenamefont {Haas}, \citenamefont {Volz}, \citenamefont {Gehr}, \citenamefont {Reichel},\ and\ \citenamefont {Estève}}]{2014Science_entangledStateFiberCavity}%
  \BibitemOpen
  \bibfield  {author} {\bibinfo {author} {\bibfnamefont {F.}~\bibnamefont {Haas}}, \bibinfo {author} {\bibfnamefont {J.}~\bibnamefont {Volz}}, \bibinfo {author} {\bibfnamefont {R.}~\bibnamefont {Gehr}}, \bibinfo {author} {\bibfnamefont {J.}~\bibnamefont {Reichel}},\ and\ \bibinfo {author} {\bibfnamefont {J.}~\bibnamefont {Estève}},\ }\bibfield  {title} {\bibinfo {title} {Entangled states of more than 40 atoms in an optical fiber cavity},\ }\href {https://doi.org/10.1126/science.1248905} {\bibfield  {journal} {\bibinfo  {journal} {Science}\ }\textbf {\bibinfo {volume} {344}},\ \bibinfo {pages} {180} (\bibinfo {year} {2014})}\BibitemShut {NoStop}%
\bibitem [{\citenamefont {Barontini}\ \emph {et~al.}(2015)\citenamefont {Barontini}, \citenamefont {Hohmann}, \citenamefont {Haas}, \citenamefont {Estève},\ and\ \citenamefont {Reichel}}]{2015Science_entanglementgeneration}%
  \BibitemOpen
  \bibfield  {author} {\bibinfo {author} {\bibfnamefont {G.}~\bibnamefont {Barontini}}, \bibinfo {author} {\bibfnamefont {L.}~\bibnamefont {Hohmann}}, \bibinfo {author} {\bibfnamefont {F.}~\bibnamefont {Haas}}, \bibinfo {author} {\bibfnamefont {J.}~\bibnamefont {Estève}},\ and\ \bibinfo {author} {\bibfnamefont {J.}~\bibnamefont {Reichel}},\ }\bibfield  {title} {\bibinfo {title} {Deterministic generation of multiparticle entanglement by quantum {Z}eno dynamics},\ }\href {https://doi.org/10.1126/science.aaa0754} {\bibfield  {journal} {\bibinfo  {journal} {Science}\ }\textbf {\bibinfo {volume} {349}},\ \bibinfo {pages} {1317} (\bibinfo {year} {2015})}\BibitemShut {NoStop}%
\bibitem [{\citenamefont {Corzo}\ \emph {et~al.}(2019)\citenamefont {Corzo}, \citenamefont {Raskop}, \citenamefont {Chandra}, \citenamefont {Sheremet}, \citenamefont {Gouraud},\ and\ \citenamefont {Laurat}}]{2019Nature_Laurat}%
  \BibitemOpen
  \bibfield  {author} {\bibinfo {author} {\bibfnamefont {N.~V.}\ \bibnamefont {Corzo}}, \bibinfo {author} {\bibfnamefont {J.}~\bibnamefont {Raskop}}, \bibinfo {author} {\bibfnamefont {A.}~\bibnamefont {Chandra}}, \bibinfo {author} {\bibfnamefont {A.~S.}\ \bibnamefont {Sheremet}}, \bibinfo {author} {\bibfnamefont {B.}~\bibnamefont {Gouraud}},\ and\ \bibinfo {author} {\bibfnamefont {J.}~\bibnamefont {Laurat}},\ }\bibfield  {title} {\bibinfo {title} {Waveguide-coupled single collective excitation of atomic arrays},\ }\href {https://doi.org/10.1038/s41586-019-0902-3} {\bibfield  {journal} {\bibinfo  {journal} {Nature}\ }\textbf {\bibinfo {volume} {566}},\ \bibinfo {pages} {359} (\bibinfo {year} {2019})}\BibitemShut {NoStop}%
\bibitem [{\citenamefont {Welte}\ \emph {et~al.}(2018)\citenamefont {Welte}, \citenamefont {Hacker}, \citenamefont {Daiss}, \citenamefont {Ritter},\ and\ \citenamefont {Rempe}}]{2018PRL_entanglement2atoms}%
  \BibitemOpen
  \bibfield  {author} {\bibinfo {author} {\bibfnamefont {S.}~\bibnamefont {Welte}}, \bibinfo {author} {\bibfnamefont {B.}~\bibnamefont {Hacker}}, \bibinfo {author} {\bibfnamefont {S.}~\bibnamefont {Daiss}}, \bibinfo {author} {\bibfnamefont {S.}~\bibnamefont {Ritter}},\ and\ \bibinfo {author} {\bibfnamefont {G.}~\bibnamefont {Rempe}},\ }\bibfield  {title} {\bibinfo {title} {Photon-mediated quantum gate between two neutral atoms in an optical cavity},\ }\href {https://doi.org/10.1103/PhysRevX.8.011018} {\bibfield  {journal} {\bibinfo  {journal} {Phys. Rev. X}\ }\textbf {\bibinfo {volume} {8}},\ \bibinfo {pages} {011018} (\bibinfo {year} {2018})}\BibitemShut {NoStop}%
\bibitem [{\citenamefont {Ðorđević}\ \emph {et~al.}(2021)\citenamefont {Ðorđević}, \citenamefont {Samutpraphoot}, \citenamefont {Ocola}, \citenamefont {Bernien}, \citenamefont {Grinkemeyer}, \citenamefont {Dimitrova}, \citenamefont {Vuletić},\ and\ \citenamefont {Lukin}}]{2022Sciecn_entanglementtransport}%
  \BibitemOpen
  \bibfield  {author} {\bibinfo {author} {\bibfnamefont {T.}~\bibnamefont {Ðorđević}}, \bibinfo {author} {\bibfnamefont {P.}~\bibnamefont {Samutpraphoot}}, \bibinfo {author} {\bibfnamefont {P.~L.}\ \bibnamefont {Ocola}}, \bibinfo {author} {\bibfnamefont {H.}~\bibnamefont {Bernien}}, \bibinfo {author} {\bibfnamefont {B.}~\bibnamefont {Grinkemeyer}}, \bibinfo {author} {\bibfnamefont {I.}~\bibnamefont {Dimitrova}}, \bibinfo {author} {\bibfnamefont {V.}~\bibnamefont {Vuletić}},\ and\ \bibinfo {author} {\bibfnamefont {M.~D.}\ \bibnamefont {Lukin}},\ }\bibfield  {title} {\bibinfo {title} {Entanglement transport and a nanophotonic interface for atoms in optical tweezers},\ }\href {https://doi.org/10.1126/science.abi9917} {\bibfield  {journal} {\bibinfo  {journal} {Science}\ }\textbf {\bibinfo {volume} {373}},\ \bibinfo {pages} {1511} (\bibinfo {year} {2021})}\BibitemShut {NoStop}%
\bibitem [{\citenamefont {Welte}\ \emph {et~al.}(2021)\citenamefont {Welte}, \citenamefont {Thomas}, \citenamefont {Hartung}, \citenamefont {Daiss}, \citenamefont {Langenfeld}, \citenamefont {Morin}, \citenamefont {Rempe},\ and\ \citenamefont {Distante}}]{2021NaturePhotonics_Rempe}%
  \BibitemOpen
  \bibfield  {author} {\bibinfo {author} {\bibfnamefont {S.}~\bibnamefont {Welte}}, \bibinfo {author} {\bibfnamefont {P.}~\bibnamefont {Thomas}}, \bibinfo {author} {\bibfnamefont {L.}~\bibnamefont {Hartung}}, \bibinfo {author} {\bibfnamefont {S.}~\bibnamefont {Daiss}}, \bibinfo {author} {\bibfnamefont {S.}~\bibnamefont {Langenfeld}}, \bibinfo {author} {\bibfnamefont {O.}~\bibnamefont {Morin}}, \bibinfo {author} {\bibfnamefont {G.}~\bibnamefont {Rempe}},\ and\ \bibinfo {author} {\bibfnamefont {E.}~\bibnamefont {Distante}},\ }\bibfield  {title} {\bibinfo {title} {A nondestructive {B}ell-state measurement on two distant atomic qubits},\ }\href {https://doi.org/10.1038/s41566-021-00802-1} {\bibfield  {journal} {\bibinfo  {journal} {Nature Photonics}\ }\textbf {\bibinfo {volume} {15}},\ \bibinfo {pages} {504} (\bibinfo {year} {2021})}\BibitemShut {NoStop}%
\bibitem [{\citenamefont {B\'eguin}\ \emph {et~al.}(2018)\citenamefont {B\'eguin}, \citenamefont {M\"uller}, \citenamefont {Appel},\ and\ \citenamefont {Polzik}}]{2018PRX_squeezingNanofiber}%
  \BibitemOpen
  \bibfield  {author} {\bibinfo {author} {\bibfnamefont {J.-B.}\ \bibnamefont {B\'eguin}}, \bibinfo {author} {\bibfnamefont {J.~H.}\ \bibnamefont {M\"uller}}, \bibinfo {author} {\bibfnamefont {J.}~\bibnamefont {Appel}},\ and\ \bibinfo {author} {\bibfnamefont {E.~S.}\ \bibnamefont {Polzik}},\ }\bibfield  {title} {\bibinfo {title} {Observation of quantum spin noise in a 1{D} light-atoms quantum interface},\ }\href {https://doi.org/10.1103/PhysRevX.8.031010} {\bibfield  {journal} {\bibinfo  {journal} {Phys. Rev. X}\ }\textbf {\bibinfo {volume} {8}},\ \bibinfo {pages} {031010} (\bibinfo {year} {2018})}\BibitemShut {NoStop}%
\bibitem [{\citenamefont {Leroux}\ \emph {et~al.}(2010)\citenamefont {Leroux}, \citenamefont {Schleier-Smith},\ and\ \citenamefont {Vuleti\ifmmode~\acute{c}\else \'{c}\fi{}}}]{2010PRL_CavitySqueezing}%
  \BibitemOpen
  \bibfield  {author} {\bibinfo {author} {\bibfnamefont {I.~D.}\ \bibnamefont {Leroux}}, \bibinfo {author} {\bibfnamefont {M.~H.}\ \bibnamefont {Schleier-Smith}},\ and\ \bibinfo {author} {\bibfnamefont {V.}~\bibnamefont {Vuleti\ifmmode~\acute{c}\else \'{c}\fi{}}},\ }\bibfield  {title} {\bibinfo {title} {Implementation of cavity squeezing of a collective atomic spin},\ }\href {https://doi.org/10.1103/PhysRevLett.104.073602} {\bibfield  {journal} {\bibinfo  {journal} {Phys. Rev. Lett.}\ }\textbf {\bibinfo {volume} {104}},\ \bibinfo {pages} {073602} (\bibinfo {year} {2010})}\BibitemShut {NoStop}%
\bibitem [{\citenamefont {Lee}\ \emph {et~al.}(2022)\citenamefont {Lee}, \citenamefont {Ding}, \citenamefont {Christensen}, \citenamefont {Rosenthal}, \citenamefont {Ison}, \citenamefont {Gillund}, \citenamefont {Bossert}, \citenamefont {Fuerschbach}, \citenamefont {Kindel}, \citenamefont {Finnegan}, \citenamefont {Wendt}, \citenamefont {Gehl}, \citenamefont {Kodigala}, \citenamefont {McGuinness}, \citenamefont {Walker}, \citenamefont {Kemme}, \citenamefont {Lentine}, \citenamefont {Biedermann},\ and\ \citenamefont {Schwindt}}]{2022NC_Lee}%
  \BibitemOpen
  \bibfield  {author} {\bibinfo {author} {\bibfnamefont {J.}~\bibnamefont {Lee}}, \bibinfo {author} {\bibfnamefont {R.}~\bibnamefont {Ding}}, \bibinfo {author} {\bibfnamefont {J.}~\bibnamefont {Christensen}}, \bibinfo {author} {\bibfnamefont {R.~R.}\ \bibnamefont {Rosenthal}}, \bibinfo {author} {\bibfnamefont {A.}~\bibnamefont {Ison}}, \bibinfo {author} {\bibfnamefont {D.~P.}\ \bibnamefont {Gillund}}, \bibinfo {author} {\bibfnamefont {D.}~\bibnamefont {Bossert}}, \bibinfo {author} {\bibfnamefont {K.~H.}\ \bibnamefont {Fuerschbach}}, \bibinfo {author} {\bibfnamefont {W.}~\bibnamefont {Kindel}}, \bibinfo {author} {\bibfnamefont {P.~S.}\ \bibnamefont {Finnegan}}, \bibinfo {author} {\bibfnamefont {J.~R.}\ \bibnamefont {Wendt}}, \bibinfo {author} {\bibfnamefont {M.}~\bibnamefont {Gehl}}, \bibinfo {author} {\bibfnamefont {A.}~\bibnamefont {Kodigala}}, \bibinfo {author} {\bibfnamefont {H.}~\bibnamefont {McGuinness}}, \bibinfo {author} {\bibfnamefont {C.~A.}\ \bibnamefont {Walker}}, \bibinfo {author} {\bibfnamefont
  {S.~A.}\ \bibnamefont {Kemme}}, \bibinfo {author} {\bibfnamefont {A.}~\bibnamefont {Lentine}}, \bibinfo {author} {\bibfnamefont {G.}~\bibnamefont {Biedermann}},\ and\ \bibinfo {author} {\bibfnamefont {P.~D.~D.}\ \bibnamefont {Schwindt}},\ }\bibfield  {title} {\bibinfo {title} {A compact cold-atom interferometer with a high data-rate grating magneto-optical trap and a photonic-integrated-circuit-compatible laser system},\ }\href {https://doi.org/10.1038/s41467-022-31410-4} {\bibfield  {journal} {\bibinfo  {journal} {Nature Communications}\ }\textbf {\bibinfo {volume} {13}},\ \bibinfo {pages} {5131} (\bibinfo {year} {2022})}\BibitemShut {NoStop}%
\bibitem [{\citenamefont {Ovchinnikov}(2022)}]{2022APL_Yuri}%
  \BibitemOpen
  \bibfield  {author} {\bibinfo {author} {\bibfnamefont {Y.~B.}\ \bibnamefont {Ovchinnikov}},\ }\bibfield  {title} {\bibinfo {title} {{A perspective on integrated atomo-photonic waveguide circuits}},\ }\href {https://doi.org/10.1063/5.0069334} {\bibfield  {journal} {\bibinfo  {journal} {Applied Physics Letters}\ }\textbf {\bibinfo {volume} {120}},\ \bibinfo {pages} {010502} (\bibinfo {year} {2022})}\BibitemShut {NoStop}%
\bibitem [{\citenamefont {Newman}\ \emph {et~al.}(2019)\citenamefont {Newman}, \citenamefont {Maurice}, \citenamefont {Drake}, \citenamefont {Stone}, \citenamefont {Briles}, \citenamefont {Spencer}, \citenamefont {Fredrick}, \citenamefont {Li}, \citenamefont {Westly}, \citenamefont {Ilic}, \citenamefont {Shen}, \citenamefont {Suh}, \citenamefont {Yang}, \citenamefont {Johnson}, \citenamefont {Johnson}, \citenamefont {Hollberg}, \citenamefont {Vahala}, \citenamefont {Srinivasan}, \citenamefont {Diddams}, \citenamefont {Kitching}, \citenamefont {Papp},\ and\ \citenamefont {Hummon}}]{2019Optica_photonicIntegrationClock}%
  \BibitemOpen
  \bibfield  {author} {\bibinfo {author} {\bibfnamefont {Z.~L.}\ \bibnamefont {Newman}}, \bibinfo {author} {\bibfnamefont {V.}~\bibnamefont {Maurice}}, \bibinfo {author} {\bibfnamefont {T.}~\bibnamefont {Drake}}, \bibinfo {author} {\bibfnamefont {J.~R.}\ \bibnamefont {Stone}}, \bibinfo {author} {\bibfnamefont {T.~C.}\ \bibnamefont {Briles}}, \bibinfo {author} {\bibfnamefont {D.~T.}\ \bibnamefont {Spencer}}, \bibinfo {author} {\bibfnamefont {C.}~\bibnamefont {Fredrick}}, \bibinfo {author} {\bibfnamefont {Q.}~\bibnamefont {Li}}, \bibinfo {author} {\bibfnamefont {D.}~\bibnamefont {Westly}}, \bibinfo {author} {\bibfnamefont {B.~R.}\ \bibnamefont {Ilic}}, \bibinfo {author} {\bibfnamefont {B.}~\bibnamefont {Shen}}, \bibinfo {author} {\bibfnamefont {M.-G.}\ \bibnamefont {Suh}}, \bibinfo {author} {\bibfnamefont {K.~Y.}\ \bibnamefont {Yang}}, \bibinfo {author} {\bibfnamefont {C.}~\bibnamefont {Johnson}}, \bibinfo {author} {\bibfnamefont {D.~M.~S.}\ \bibnamefont {Johnson}}, \bibinfo {author} {\bibfnamefont
  {L.}~\bibnamefont {Hollberg}}, \bibinfo {author} {\bibfnamefont {K.~J.}\ \bibnamefont {Vahala}}, \bibinfo {author} {\bibfnamefont {K.}~\bibnamefont {Srinivasan}}, \bibinfo {author} {\bibfnamefont {S.~A.}\ \bibnamefont {Diddams}}, \bibinfo {author} {\bibfnamefont {J.}~\bibnamefont {Kitching}}, \bibinfo {author} {\bibfnamefont {S.~B.}\ \bibnamefont {Papp}},\ and\ \bibinfo {author} {\bibfnamefont {M.~T.}\ \bibnamefont {Hummon}},\ }\bibfield  {title} {\bibinfo {title} {Architecture for the photonic integration of an optical atomic clock},\ }\href {https://doi.org/10.1364/OPTICA.6.000680} {\bibfield  {journal} {\bibinfo  {journal} {Optica}\ }\textbf {\bibinfo {volume} {6}},\ \bibinfo {pages} {680} (\bibinfo {year} {2019})}\BibitemShut {NoStop}%
\bibitem [{\citenamefont {Pedrozo-Pe{\~{n}}afiel}\ \emph {et~al.}(2020)\citenamefont {Pedrozo-Pe{\~{n}}afiel}, \citenamefont {Colombo}, \citenamefont {Shu}, \citenamefont {Adiyatullin}, \citenamefont {Li}, \citenamefont {Mendez}, \citenamefont {Braverman}, \citenamefont {Kawasaki}, \citenamefont {Akamatsu}, \citenamefont {Xiao},\ and\ \citenamefont {Vuleti{\'{c}}}}]{2020Nature_EntanglementClock}%
  \BibitemOpen
  \bibfield  {author} {\bibinfo {author} {\bibfnamefont {E.}~\bibnamefont {Pedrozo-Pe{\~{n}}afiel}}, \bibinfo {author} {\bibfnamefont {S.}~\bibnamefont {Colombo}}, \bibinfo {author} {\bibfnamefont {C.}~\bibnamefont {Shu}}, \bibinfo {author} {\bibfnamefont {A.~F.}\ \bibnamefont {Adiyatullin}}, \bibinfo {author} {\bibfnamefont {Z.}~\bibnamefont {Li}}, \bibinfo {author} {\bibfnamefont {E.}~\bibnamefont {Mendez}}, \bibinfo {author} {\bibfnamefont {B.}~\bibnamefont {Braverman}}, \bibinfo {author} {\bibfnamefont {A.}~\bibnamefont {Kawasaki}}, \bibinfo {author} {\bibfnamefont {D.}~\bibnamefont {Akamatsu}}, \bibinfo {author} {\bibfnamefont {Y.}~\bibnamefont {Xiao}},\ and\ \bibinfo {author} {\bibfnamefont {V.}~\bibnamefont {Vuleti{\'{c}}}},\ }\bibfield  {title} {\bibinfo {title} {Entanglement on an optical atomic-clock transition},\ }\href {https://doi.org/10.1038/s41586-020-3006-1} {\bibfield  {journal} {\bibinfo  {journal} {Nature}\ }\textbf {\bibinfo {volume} {588}},\ \bibinfo {pages} {414} (\bibinfo {year}
  {2020})}\BibitemShut {NoStop}%
\bibitem [{\citenamefont {K\'om\'ar}\ \emph {et~al.}(2016)\citenamefont {K\'om\'ar}, \citenamefont {Topcu}, \citenamefont {Kessler}, \citenamefont {Derevianko}, \citenamefont {Vuleti\ifmmode~\acute{c}\else \'{c}\fi{}}, \citenamefont {Ye},\ and\ \citenamefont {Lukin}}]{2016PRL_networkofClock}%
  \BibitemOpen
  \bibfield  {author} {\bibinfo {author} {\bibfnamefont {P.}~\bibnamefont {K\'om\'ar}}, \bibinfo {author} {\bibfnamefont {T.}~\bibnamefont {Topcu}}, \bibinfo {author} {\bibfnamefont {E.~M.}\ \bibnamefont {Kessler}}, \bibinfo {author} {\bibfnamefont {A.}~\bibnamefont {Derevianko}}, \bibinfo {author} {\bibfnamefont {V.}~\bibnamefont {Vuleti\ifmmode~\acute{c}\else \'{c}\fi{}}}, \bibinfo {author} {\bibfnamefont {J.}~\bibnamefont {Ye}},\ and\ \bibinfo {author} {\bibfnamefont {M.~D.}\ \bibnamefont {Lukin}},\ }\bibfield  {title} {\bibinfo {title} {Quantum network of atom clocks: A possible implementation with neutral atoms},\ }\href {https://doi.org/10.1103/PhysRevLett.117.060506} {\bibfield  {journal} {\bibinfo  {journal} {Phys. Rev. Lett.}\ }\textbf {\bibinfo {volume} {117}},\ \bibinfo {pages} {060506} (\bibinfo {year} {2016})}\BibitemShut {NoStop}%
\bibitem [{\citenamefont {Pérez-Ríos}\ \emph {et~al.}(2017)\citenamefont {Pérez-Ríos}, \citenamefont {Kim},\ and\ \citenamefont {Hung}}]{2017NJP_molecule}%
  \BibitemOpen
  \bibfield  {author} {\bibinfo {author} {\bibfnamefont {J.}~\bibnamefont {Pérez-Ríos}}, \bibinfo {author} {\bibfnamefont {M.~E.}\ \bibnamefont {Kim}},\ and\ \bibinfo {author} {\bibfnamefont {C.-L.}\ \bibnamefont {Hung}},\ }\bibfield  {title} {\bibinfo {title} {Ultracold molecule assembly with photonic crystals},\ }\href {https://doi.org/10.1088/1367-2630/aa9b49} {\bibfield  {journal} {\bibinfo  {journal} {New Journal of Physics}\ }\textbf {\bibinfo {volume} {19}},\ \bibinfo {pages} {123035} (\bibinfo {year} {2017})}\BibitemShut {NoStop}%
\bibitem [{\citenamefont {Kampschulte}\ and\ \citenamefont {Denschlag}(2018)}]{2018NJP_molecules}%
  \BibitemOpen
  \bibfield  {author} {\bibinfo {author} {\bibfnamefont {T.}~\bibnamefont {Kampschulte}}\ and\ \bibinfo {author} {\bibfnamefont {J.~H.}\ \bibnamefont {Denschlag}},\ }\bibfield  {title} {\bibinfo {title} {Cavity-controlled formation of ultracold molecules},\ }\href {https://doi.org/10.1088/1367-2630/aaf5f5} {\bibfield  {journal} {\bibinfo  {journal} {New Journal of Physics}\ }\textbf {\bibinfo {volume} {20}},\ \bibinfo {pages} {123015} (\bibinfo {year} {2018})}\BibitemShut {NoStop}%
\bibitem [{\citenamefont {Wellnitz}\ \emph {et~al.}(2020)\citenamefont {Wellnitz}, \citenamefont {Sch\"utz}, \citenamefont {Whitlock}, \citenamefont {Schachenmayer},\ and\ \citenamefont {Pupillo}}]{2020PRL_MoleculeFormation}%
  \BibitemOpen
  \bibfield  {author} {\bibinfo {author} {\bibfnamefont {D.}~\bibnamefont {Wellnitz}}, \bibinfo {author} {\bibfnamefont {S.}~\bibnamefont {Sch\"utz}}, \bibinfo {author} {\bibfnamefont {S.}~\bibnamefont {Whitlock}}, \bibinfo {author} {\bibfnamefont {J.}~\bibnamefont {Schachenmayer}},\ and\ \bibinfo {author} {\bibfnamefont {G.}~\bibnamefont {Pupillo}},\ }\bibfield  {title} {\bibinfo {title} {Collective dissipative molecule formation in a cavity},\ }\href {https://doi.org/10.1103/PhysRevLett.125.193201} {\bibfield  {journal} {\bibinfo  {journal} {Phys. Rev. Lett.}\ }\textbf {\bibinfo {volume} {125}},\ \bibinfo {pages} {193201} (\bibinfo {year} {2020})}\BibitemShut {NoStop}%
\bibitem [{\citenamefont {Vetsch}\ \emph {et~al.}(2010)\citenamefont {Vetsch}, \citenamefont {Reitz}, \citenamefont {Sagu\'e}, \citenamefont {Schmidt}, \citenamefont {Dawkins},\ and\ \citenamefont {Rauschenbeutel}}]{2010PRL_nanofiber}%
  \BibitemOpen
  \bibfield  {author} {\bibinfo {author} {\bibfnamefont {E.}~\bibnamefont {Vetsch}}, \bibinfo {author} {\bibfnamefont {D.}~\bibnamefont {Reitz}}, \bibinfo {author} {\bibfnamefont {G.}~\bibnamefont {Sagu\'e}}, \bibinfo {author} {\bibfnamefont {R.}~\bibnamefont {Schmidt}}, \bibinfo {author} {\bibfnamefont {S.~T.}\ \bibnamefont {Dawkins}},\ and\ \bibinfo {author} {\bibfnamefont {A.}~\bibnamefont {Rauschenbeutel}},\ }\bibfield  {title} {\bibinfo {title} {Optical interface created by laser-cooled atoms trapped in the evanescent field surrounding an optical nanofiber},\ }\href {https://doi.org/10.1103/PhysRevLett.104.203603} {\bibfield  {journal} {\bibinfo  {journal} {Phys. Rev. Lett.}\ }\textbf {\bibinfo {volume} {104}},\ \bibinfo {pages} {203603} (\bibinfo {year} {2010})}\BibitemShut {NoStop}%
\bibitem [{\citenamefont {Goban}\ \emph {et~al.}(2012)\citenamefont {Goban}, \citenamefont {Choi}, \citenamefont {Alton}, \citenamefont {Ding}, \citenamefont {Lacro\^ute}, \citenamefont {Pototschnig}, \citenamefont {Thiele}, \citenamefont {Stern},\ and\ \citenamefont {Kimble}}]{2012PRL_nanofiber}%
  \BibitemOpen
  \bibfield  {author} {\bibinfo {author} {\bibfnamefont {A.}~\bibnamefont {Goban}}, \bibinfo {author} {\bibfnamefont {K.~S.}\ \bibnamefont {Choi}}, \bibinfo {author} {\bibfnamefont {D.~J.}\ \bibnamefont {Alton}}, \bibinfo {author} {\bibfnamefont {D.}~\bibnamefont {Ding}}, \bibinfo {author} {\bibfnamefont {C.}~\bibnamefont {Lacro\^ute}}, \bibinfo {author} {\bibfnamefont {M.}~\bibnamefont {Pototschnig}}, \bibinfo {author} {\bibfnamefont {T.}~\bibnamefont {Thiele}}, \bibinfo {author} {\bibfnamefont {N.~P.}\ \bibnamefont {Stern}},\ and\ \bibinfo {author} {\bibfnamefont {H.~J.}\ \bibnamefont {Kimble}},\ }\bibfield  {title} {\bibinfo {title} {Demonstration of a state-insensitive, compensated nanofiber trap},\ }\href {https://doi.org/10.1103/PhysRevLett.109.033603} {\bibfield  {journal} {\bibinfo  {journal} {Phys. Rev. Lett.}\ }\textbf {\bibinfo {volume} {109}},\ \bibinfo {pages} {033603} (\bibinfo {year} {2012})}\BibitemShut {NoStop}%
\bibitem [{\citenamefont {Kestler}\ \emph {et~al.}(2023)\citenamefont {Kestler}, \citenamefont {Ton}, \citenamefont {Filin}, \citenamefont {Cheung}, \citenamefont {Schneeweiss}, \citenamefont {Hoinkes}, \citenamefont {Volz}, \citenamefont {Safronova}, \citenamefont {Rauschenbeutel},\ and\ \citenamefont {Barreiro}}]{2023PRXQuantum_Srnanofiber}%
  \BibitemOpen
  \bibfield  {author} {\bibinfo {author} {\bibfnamefont {G.}~\bibnamefont {Kestler}}, \bibinfo {author} {\bibfnamefont {K.}~\bibnamefont {Ton}}, \bibinfo {author} {\bibfnamefont {D.}~\bibnamefont {Filin}}, \bibinfo {author} {\bibfnamefont {C.}~\bibnamefont {Cheung}}, \bibinfo {author} {\bibfnamefont {P.}~\bibnamefont {Schneeweiss}}, \bibinfo {author} {\bibfnamefont {T.}~\bibnamefont {Hoinkes}}, \bibinfo {author} {\bibfnamefont {J.}~\bibnamefont {Volz}}, \bibinfo {author} {\bibfnamefont {M.}~\bibnamefont {Safronova}}, \bibinfo {author} {\bibfnamefont {A.}~\bibnamefont {Rauschenbeutel}},\ and\ \bibinfo {author} {\bibfnamefont {J.}~\bibnamefont {Barreiro}},\ }\bibfield  {title} {\bibinfo {title} {State-insensitive trapping of alkaline-earth atoms in a nanofiber-based optical dipole trap},\ }\href {https://doi.org/10.1103/PRXQuantum.4.040308} {\bibfield  {journal} {\bibinfo  {journal} {PRX Quantum}\ }\textbf {\bibinfo {volume} {4}},\ \bibinfo {pages} {040308} (\bibinfo {year} {2023})}\BibitemShut {NoStop}%
\bibitem [{\citenamefont {Thompson}\ \emph {et~al.}(2013)\citenamefont {Thompson}, \citenamefont {Tiecke}, \citenamefont {de~Leon}, \citenamefont {Feist}, \citenamefont {Akimov}, \citenamefont {Gullans}, \citenamefont {Zibrov}, \citenamefont {Vuletić},\ and\ \citenamefont {Lukin}}]{2013Science_tweezertrap}%
  \BibitemOpen
  \bibfield  {author} {\bibinfo {author} {\bibfnamefont {J.~D.}\ \bibnamefont {Thompson}}, \bibinfo {author} {\bibfnamefont {T.~G.}\ \bibnamefont {Tiecke}}, \bibinfo {author} {\bibfnamefont {N.~P.}\ \bibnamefont {de~Leon}}, \bibinfo {author} {\bibfnamefont {J.}~\bibnamefont {Feist}}, \bibinfo {author} {\bibfnamefont {A.~V.}\ \bibnamefont {Akimov}}, \bibinfo {author} {\bibfnamefont {M.}~\bibnamefont {Gullans}}, \bibinfo {author} {\bibfnamefont {A.~S.}\ \bibnamefont {Zibrov}}, \bibinfo {author} {\bibfnamefont {V.}~\bibnamefont {Vuletić}},\ and\ \bibinfo {author} {\bibfnamefont {M.~D.}\ \bibnamefont {Lukin}},\ }\bibfield  {title} {\bibinfo {title} {Coupling a single trapped atom to a nanoscale optical cavity},\ }\href {https://doi.org/10.1126/science.1237125} {\bibfield  {journal} {\bibinfo  {journal} {Science}\ }\textbf {\bibinfo {volume} {340}},\ \bibinfo {pages} {1202} (\bibinfo {year} {2013})}\BibitemShut {NoStop}%
\bibitem [{\citenamefont {Burgers}\ \emph {et~al.}(2019)\citenamefont {Burgers}, \citenamefont {Peng}, \citenamefont {Muniz}, \citenamefont {McClung}, \citenamefont {Martin},\ and\ \citenamefont {Kimble}}]{2019PNAS_Kimble}%
  \BibitemOpen
  \bibfield  {author} {\bibinfo {author} {\bibfnamefont {A.~P.}\ \bibnamefont {Burgers}}, \bibinfo {author} {\bibfnamefont {L.~S.}\ \bibnamefont {Peng}}, \bibinfo {author} {\bibfnamefont {J.~A.}\ \bibnamefont {Muniz}}, \bibinfo {author} {\bibfnamefont {A.~C.}\ \bibnamefont {McClung}}, \bibinfo {author} {\bibfnamefont {M.~J.}\ \bibnamefont {Martin}},\ and\ \bibinfo {author} {\bibfnamefont {H.~J.}\ \bibnamefont {Kimble}},\ }\bibfield  {title} {\bibinfo {title} {Clocked atom delivery to a photonic crystal waveguide},\ }\href {https://doi.org/10.1073/pnas.1817249115} {\bibfield  {journal} {\bibinfo  {journal} {Proceedings of the National Academy of Sciences}\ }\textbf {\bibinfo {volume} {116}},\ \bibinfo {pages} {456} (\bibinfo {year} {2019})}\BibitemShut {NoStop}%
\bibitem [{\citenamefont {Kim}\ \emph {et~al.}(2019)\citenamefont {Kim}, \citenamefont {Chang}, \citenamefont {Fields}, \citenamefont {Chen},\ and\ \citenamefont {Hung}}]{2019NC_imaging}%
  \BibitemOpen
  \bibfield  {author} {\bibinfo {author} {\bibfnamefont {M.~E.}\ \bibnamefont {Kim}}, \bibinfo {author} {\bibfnamefont {T.-H.}\ \bibnamefont {Chang}}, \bibinfo {author} {\bibfnamefont {B.~M.}\ \bibnamefont {Fields}}, \bibinfo {author} {\bibfnamefont {C.-A.}\ \bibnamefont {Chen}},\ and\ \bibinfo {author} {\bibfnamefont {C.-L.}\ \bibnamefont {Hung}},\ }\bibfield  {title} {\bibinfo {title} {Trapping single atoms on a nanophotonic circuit with configurable tweezer lattices},\ }\href {https://doi.org/10.1038/s41467-019-09635-7} {\bibfield  {journal} {\bibinfo  {journal} {Nature Communications}\ }\textbf {\bibinfo {volume} {10}},\ \bibinfo {pages} {1647} (\bibinfo {year} {2019})}\BibitemShut {NoStop}%
\bibitem [{\citenamefont {Xu}\ \emph {et~al.}(2023)\citenamefont {Xu}, \citenamefont {Wang}, \citenamefont {Chen}, \citenamefont {Chen}, \citenamefont {Yang}, \citenamefont {Xu}, \citenamefont {Liu}, \citenamefont {Li}, \citenamefont {Guo}, \citenamefont {Zou},\ and\ \citenamefont {Xiang}}]{2023CPL_conveyor-belt}%
  \BibitemOpen
  \bibfield  {author} {\bibinfo {author} {\bibfnamefont {L.}~\bibnamefont {Xu}}, \bibinfo {author} {\bibfnamefont {L.-X.}\ \bibnamefont {Wang}}, \bibinfo {author} {\bibfnamefont {G.-J.}\ \bibnamefont {Chen}}, \bibinfo {author} {\bibfnamefont {L.}~\bibnamefont {Chen}}, \bibinfo {author} {\bibfnamefont {Y.-H.}\ \bibnamefont {Yang}}, \bibinfo {author} {\bibfnamefont {X.-B.}\ \bibnamefont {Xu}}, \bibinfo {author} {\bibfnamefont {A.}~\bibnamefont {Liu}}, \bibinfo {author} {\bibfnamefont {C.-F.}\ \bibnamefont {Li}}, \bibinfo {author} {\bibfnamefont {G.-C.}\ \bibnamefont {Guo}}, \bibinfo {author} {\bibfnamefont {C.-L.}\ \bibnamefont {Zou}},\ and\ \bibinfo {author} {\bibfnamefont {G.-Y.}\ \bibnamefont {Xiang}},\ }\bibfield  {title} {\bibinfo {title} {Transporting cold atoms towards a gan-on-sapphire chip via an optical conveyor belt},\ }\href {https://doi.org/10.1088/0256-307X/40/9/093701} {\bibfield  {journal} {\bibinfo  {journal} {Chinese Physics Letters}\ }\textbf {\bibinfo {volume} {40}},\ \bibinfo {pages}
  {093701} (\bibinfo {year} {2023})}\BibitemShut {NoStop}%
\bibitem [{\citenamefont {Zhou}\ \emph {et~al.}(2023)\citenamefont {Zhou}, \citenamefont {Tamura}, \citenamefont {Chang},\ and\ \citenamefont {Hung}}]{2023PRL_Coupling}%
  \BibitemOpen
  \bibfield  {author} {\bibinfo {author} {\bibfnamefont {X.}~\bibnamefont {Zhou}}, \bibinfo {author} {\bibfnamefont {H.}~\bibnamefont {Tamura}}, \bibinfo {author} {\bibfnamefont {T.-H.}\ \bibnamefont {Chang}},\ and\ \bibinfo {author} {\bibfnamefont {C.-L.}\ \bibnamefont {Hung}},\ }\bibfield  {title} {\bibinfo {title} {Coupling single atoms to a nanophotonic whispering-gallery-mode resonator via optical guiding},\ }\href {https://doi.org/10.1103/PhysRevLett.130.103601} {\bibfield  {journal} {\bibinfo  {journal} {Phys. Rev. Lett.}\ }\textbf {\bibinfo {volume} {130}},\ \bibinfo {pages} {103601} (\bibinfo {year} {2023})}\BibitemShut {NoStop}%
\bibitem [{\citenamefont {Vuleti\ifmmode~\acute{c}\else \'{c}\fi{}}\ \emph {et~al.}(1998)\citenamefont {Vuleti\ifmmode~\acute{c}\else \'{c}\fi{}}, \citenamefont {Chin}, \citenamefont {Kerman},\ and\ \citenamefont {Chu}}]{1998PRL_dRSC}%
  \BibitemOpen
  \bibfield  {author} {\bibinfo {author} {\bibfnamefont {V.}~\bibnamefont {Vuleti\ifmmode~\acute{c}\else \'{c}\fi{}}}, \bibinfo {author} {\bibfnamefont {C.}~\bibnamefont {Chin}}, \bibinfo {author} {\bibfnamefont {A.~J.}\ \bibnamefont {Kerman}},\ and\ \bibinfo {author} {\bibfnamefont {S.}~\bibnamefont {Chu}},\ }\bibfield  {title} {\bibinfo {title} {Degenerate {R}aman sideband cooling of trapped cesium atoms at very high atomic densities},\ }\href {https://doi.org/10.1103/PhysRevLett.81.5768} {\bibfield  {journal} {\bibinfo  {journal} {Phys. Rev. Lett.}\ }\textbf {\bibinfo {volume} {81}},\ \bibinfo {pages} {5768} (\bibinfo {year} {1998})}\BibitemShut {NoStop}%
\bibitem [{\citenamefont {Kerman}\ \emph {et~al.}(2000)\citenamefont {Kerman}, \citenamefont {Vuleti\ifmmode~\acute{c}\else \'{c}\fi{}}, \citenamefont {Chin},\ and\ \citenamefont {Chu}}]{2000PRL_dRSC}%
  \BibitemOpen
  \bibfield  {author} {\bibinfo {author} {\bibfnamefont {A.~J.}\ \bibnamefont {Kerman}}, \bibinfo {author} {\bibfnamefont {V.}~\bibnamefont {Vuleti\ifmmode~\acute{c}\else \'{c}\fi{}}}, \bibinfo {author} {\bibfnamefont {C.}~\bibnamefont {Chin}},\ and\ \bibinfo {author} {\bibfnamefont {S.}~\bibnamefont {Chu}},\ }\bibfield  {title} {\bibinfo {title} {Beyond optical molasses: 3{D} {R}aman sideband cooling of atomic cesium to high phase-space density},\ }\href {https://doi.org/10.1103/PhysRevLett.84.439} {\bibfield  {journal} {\bibinfo  {journal} {Phys. Rev. Lett.}\ }\textbf {\bibinfo {volume} {84}},\ \bibinfo {pages} {439} (\bibinfo {year} {2000})}\BibitemShut {NoStop}%
\bibitem [{\citenamefont {Mechelen}\ and\ \citenamefont {Jacob}(2016)}]{2016Optica_evanescent}%
  \BibitemOpen
  \bibfield  {author} {\bibinfo {author} {\bibfnamefont {T.~V.}\ \bibnamefont {Mechelen}}\ and\ \bibinfo {author} {\bibfnamefont {Z.}~\bibnamefont {Jacob}},\ }\bibfield  {title} {\bibinfo {title} {Universal spin-momentum locking of evanescent waves},\ }\href {https://doi.org/10.1364/OPTICA.3.000118} {\bibfield  {journal} {\bibinfo  {journal} {Optica}\ }\textbf {\bibinfo {volume} {3}},\ \bibinfo {pages} {118} (\bibinfo {year} {2016})}\BibitemShut {NoStop}%
\bibitem [{\citenamefont {Chang}\ \emph {et~al.}(2019)\citenamefont {Chang}, \citenamefont {Fields}, \citenamefont {Kim},\ and\ \citenamefont {Hung}}]{2019Optica_Chang}%
  \BibitemOpen
  \bibfield  {author} {\bibinfo {author} {\bibfnamefont {T.-H.}\ \bibnamefont {Chang}}, \bibinfo {author} {\bibfnamefont {B.~M.}\ \bibnamefont {Fields}}, \bibinfo {author} {\bibfnamefont {M.~E.}\ \bibnamefont {Kim}},\ and\ \bibinfo {author} {\bibfnamefont {C.-L.}\ \bibnamefont {Hung}},\ }\bibfield  {title} {\bibinfo {title} {Microring resonators on a suspended membrane circuit for atom-light interactions},\ }\href {https://doi.org/10.1364/OPTICA.6.001203} {\bibfield  {journal} {\bibinfo  {journal} {Optica}\ }\textbf {\bibinfo {volume} {6}},\ \bibinfo {pages} {1203} (\bibinfo {year} {2019})}\BibitemShut {NoStop}%
\bibitem [{\citenamefont {Lodahl}\ \emph {et~al.}(2017)\citenamefont {Lodahl}, \citenamefont {Mahmoodian}, \citenamefont {Stobbe}, \citenamefont {Rauschenbeutel}, \citenamefont {Schneeweiss}, \citenamefont {Volz}, \citenamefont {Pichler},\ and\ \citenamefont {Zoller}}]{2017Nature_ChiralQuantumOptics}%
  \BibitemOpen
  \bibfield  {author} {\bibinfo {author} {\bibfnamefont {P.}~\bibnamefont {Lodahl}}, \bibinfo {author} {\bibfnamefont {S.}~\bibnamefont {Mahmoodian}}, \bibinfo {author} {\bibfnamefont {S.}~\bibnamefont {Stobbe}}, \bibinfo {author} {\bibfnamefont {A.}~\bibnamefont {Rauschenbeutel}}, \bibinfo {author} {\bibfnamefont {P.}~\bibnamefont {Schneeweiss}}, \bibinfo {author} {\bibfnamefont {J.}~\bibnamefont {Volz}}, \bibinfo {author} {\bibfnamefont {H.}~\bibnamefont {Pichler}},\ and\ \bibinfo {author} {\bibfnamefont {P.}~\bibnamefont {Zoller}},\ }\bibfield  {title} {\bibinfo {title} {Chiral quantum optics},\ }\href {https://doi.org/10.1038/nature21037} {\bibfield  {journal} {\bibinfo  {journal} {Nature}\ }\textbf {\bibinfo {volume} {541}},\ \bibinfo {pages} {473} (\bibinfo {year} {2017})}\BibitemShut {NoStop}%
\bibitem [{\citenamefont {Jennewein}\ \emph {et~al.}(2016)\citenamefont {Jennewein}, \citenamefont {Besbes}, \citenamefont {Schilder}, \citenamefont {Jenkins}, \citenamefont {Sauvan}, \citenamefont {Ruostekoski}, \citenamefont {Greffet}, \citenamefont {Sortais},\ and\ \citenamefont {Browaeys}}]{2016PRL_Browaeys_shifts}%
  \BibitemOpen
  \bibfield  {author} {\bibinfo {author} {\bibfnamefont {S.}~\bibnamefont {Jennewein}}, \bibinfo {author} {\bibfnamefont {M.}~\bibnamefont {Besbes}}, \bibinfo {author} {\bibfnamefont {N.~J.}\ \bibnamefont {Schilder}}, \bibinfo {author} {\bibfnamefont {S.~D.}\ \bibnamefont {Jenkins}}, \bibinfo {author} {\bibfnamefont {C.}~\bibnamefont {Sauvan}}, \bibinfo {author} {\bibfnamefont {J.}~\bibnamefont {Ruostekoski}}, \bibinfo {author} {\bibfnamefont {J.-J.}\ \bibnamefont {Greffet}}, \bibinfo {author} {\bibfnamefont {Y.~R.~P.}\ \bibnamefont {Sortais}},\ and\ \bibinfo {author} {\bibfnamefont {A.}~\bibnamefont {Browaeys}},\ }\bibfield  {title} {\bibinfo {title} {Coherent scattering of near-resonant light by a dense microscopic cold atomic cloud},\ }\href {https://doi.org/10.1103/PhysRevLett.116.233601} {\bibfield  {journal} {\bibinfo  {journal} {Phys. Rev. Lett.}\ }\textbf {\bibinfo {volume} {116}},\ \bibinfo {pages} {233601} (\bibinfo {year} {2016})}\BibitemShut {NoStop}%
\bibitem [{\citenamefont {Sutherland}\ and\ \citenamefont {Robicheaux}(2016)}]{2016PRA_Francis_atomicarray}%
  \BibitemOpen
  \bibfield  {author} {\bibinfo {author} {\bibfnamefont {R.~T.}\ \bibnamefont {Sutherland}}\ and\ \bibinfo {author} {\bibfnamefont {F.}~\bibnamefont {Robicheaux}},\ }\bibfield  {title} {\bibinfo {title} {Collective dipole-dipole interactions in an atomic array},\ }\href {https://doi.org/10.1103/PhysRevA.94.013847} {\bibfield  {journal} {\bibinfo  {journal} {Phys. Rev. A}\ }\textbf {\bibinfo {volume} {94}},\ \bibinfo {pages} {013847} (\bibinfo {year} {2016})}\BibitemShut {NoStop}%
\bibitem [{\citenamefont {Scully}(2009)}]{2009PRL_Scully}%
  \BibitemOpen
  \bibfield  {author} {\bibinfo {author} {\bibfnamefont {M.~O.}\ \bibnamefont {Scully}},\ }\bibfield  {title} {\bibinfo {title} {Collective {L}amb shift in single photon {D}icke superradiance},\ }\href {https://doi.org/10.1103/PhysRevLett.102.143601} {\bibfield  {journal} {\bibinfo  {journal} {Phys. Rev. Lett.}\ }\textbf {\bibinfo {volume} {102}},\ \bibinfo {pages} {143601} (\bibinfo {year} {2009})}\BibitemShut {NoStop}%
\bibitem [{\citenamefont {Meng}\ \emph {et~al.}(2018)\citenamefont {Meng}, \citenamefont {Dareau}, \citenamefont {Schneeweiss},\ and\ \citenamefont {Rauschenbeutel}}]{2018PRX_Nanofibercooling}%
  \BibitemOpen
  \bibfield  {author} {\bibinfo {author} {\bibfnamefont {Y.}~\bibnamefont {Meng}}, \bibinfo {author} {\bibfnamefont {A.}~\bibnamefont {Dareau}}, \bibinfo {author} {\bibfnamefont {P.}~\bibnamefont {Schneeweiss}},\ and\ \bibinfo {author} {\bibfnamefont {A.}~\bibnamefont {Rauschenbeutel}},\ }\bibfield  {title} {\bibinfo {title} {Near-ground-state cooling of atoms optically trapped 300 nm away from a hot surface},\ }\href {https://doi.org/10.1103/PhysRevX.8.031054} {\bibfield  {journal} {\bibinfo  {journal} {Phys. Rev. X}\ }\textbf {\bibinfo {volume} {8}},\ \bibinfo {pages} {031054} (\bibinfo {year} {2018})}\BibitemShut {NoStop}%
\bibitem [{\citenamefont {Hu}\ \emph {et~al.}(2017)\citenamefont {Hu}, \citenamefont {Urvoy}, \citenamefont {Vendeiro}, \citenamefont {Crépel}, \citenamefont {Chen},\ and\ \citenamefont {Vuletić}}]{2017Science_Vladan}%
  \BibitemOpen
  \bibfield  {author} {\bibinfo {author} {\bibfnamefont {J.}~\bibnamefont {Hu}}, \bibinfo {author} {\bibfnamefont {A.}~\bibnamefont {Urvoy}}, \bibinfo {author} {\bibfnamefont {Z.}~\bibnamefont {Vendeiro}}, \bibinfo {author} {\bibfnamefont {V.}~\bibnamefont {Crépel}}, \bibinfo {author} {\bibfnamefont {W.}~\bibnamefont {Chen}},\ and\ \bibinfo {author} {\bibfnamefont {V.}~\bibnamefont {Vuletić}},\ }\bibfield  {title} {\bibinfo {title} {Creation of a {B}ose-condensed gas of {$^{87}$}{Rb} by laser cooling},\ }\href {https://doi.org/10.1126/science.aan5614} {\bibfield  {journal} {\bibinfo  {journal} {Science}\ }\textbf {\bibinfo {volume} {358}},\ \bibinfo {pages} {1078} (\bibinfo {year} {2017})}\BibitemShut {NoStop}%
\bibitem [{\citenamefont {Urvoy}\ \emph {et~al.}(2019)\citenamefont {Urvoy}, \citenamefont {Vendeiro}, \citenamefont {Ramette}, \citenamefont {Adiyatullin},\ and\ \citenamefont {Vuleti\ifmmode~\acute{c}\else \'{c}\fi{}}}]{2019PRL_Rb_Vladan}%
  \BibitemOpen
  \bibfield  {author} {\bibinfo {author} {\bibfnamefont {A.}~\bibnamefont {Urvoy}}, \bibinfo {author} {\bibfnamefont {Z.}~\bibnamefont {Vendeiro}}, \bibinfo {author} {\bibfnamefont {J.}~\bibnamefont {Ramette}}, \bibinfo {author} {\bibfnamefont {A.}~\bibnamefont {Adiyatullin}},\ and\ \bibinfo {author} {\bibfnamefont {V.}~\bibnamefont {Vuleti\ifmmode~\acute{c}\else \'{c}\fi{}}},\ }\bibfield  {title} {\bibinfo {title} {Direct laser cooling to {B}ose-{E}instein {C}ondensation in a dipole trap},\ }\href {https://doi.org/10.1103/PhysRevLett.122.203202} {\bibfield  {journal} {\bibinfo  {journal} {Phys. Rev. Lett.}\ }\textbf {\bibinfo {volume} {122}},\ \bibinfo {pages} {203202} (\bibinfo {year} {2019})}\BibitemShut {NoStop}%
\bibitem [{\citenamefont {Solano}\ \emph {et~al.}(2019)\citenamefont {Solano}, \citenamefont {Duan}, \citenamefont {Chen}, \citenamefont {Rudelis}, \citenamefont {Chin},\ and\ \citenamefont {Vuleti\ifmmode~\acute{c}\else \'{c}\fi{}}}]{2019PRL_Cesium_Vladan}%
  \BibitemOpen
  \bibfield  {author} {\bibinfo {author} {\bibfnamefont {P.}~\bibnamefont {Solano}}, \bibinfo {author} {\bibfnamefont {Y.}~\bibnamefont {Duan}}, \bibinfo {author} {\bibfnamefont {Y.-T.}\ \bibnamefont {Chen}}, \bibinfo {author} {\bibfnamefont {A.}~\bibnamefont {Rudelis}}, \bibinfo {author} {\bibfnamefont {C.}~\bibnamefont {Chin}},\ and\ \bibinfo {author} {\bibfnamefont {V.}~\bibnamefont {Vuleti\ifmmode~\acute{c}\else \'{c}\fi{}}},\ }\bibfield  {title} {\bibinfo {title} {Strongly correlated quantum gas prepared by direct laser cooling},\ }\href {https://doi.org/10.1103/PhysRevLett.123.173401} {\bibfield  {journal} {\bibinfo  {journal} {Phys. Rev. Lett.}\ }\textbf {\bibinfo {volume} {123}},\ \bibinfo {pages} {173401} (\bibinfo {year} {2019})}\BibitemShut {NoStop}%
\bibitem [{\citenamefont {Chang}\ \emph {et~al.}(2022)\citenamefont {Chang}, \citenamefont {Zhou}, \citenamefont {Tamura},\ and\ \citenamefont {Hung}}]{2022OE_3Dcoupler}%
  \BibitemOpen
  \bibfield  {author} {\bibinfo {author} {\bibfnamefont {T.-H.}\ \bibnamefont {Chang}}, \bibinfo {author} {\bibfnamefont {X.}~\bibnamefont {Zhou}}, \bibinfo {author} {\bibfnamefont {H.}~\bibnamefont {Tamura}},\ and\ \bibinfo {author} {\bibfnamefont {C.-L.}\ \bibnamefont {Hung}},\ }\bibfield  {title} {\bibinfo {title} {Realization of efficient 3{D} tapered waveguide-to-fiber couplers on a nanophotonic circuit},\ }\href {https://doi.org/10.1364/OE.468738} {\bibfield  {journal} {\bibinfo  {journal} {Opt. Express}\ }\textbf {\bibinfo {volume} {30}},\ \bibinfo {pages} {31643} (\bibinfo {year} {2022})}\BibitemShut {NoStop}%
\bibitem [{\citenamefont {S\"oding}\ \emph {et~al.}(1998)\citenamefont {S\"oding}, \citenamefont {Gu\'ery-Odelin}, \citenamefont {Desbiolles}, \citenamefont {Ferrari},\ and\ \citenamefont {Dalibard}}]{soding1998giant}%
  \BibitemOpen
  \bibfield  {author} {\bibinfo {author} {\bibfnamefont {J.}~\bibnamefont {S\"oding}}, \bibinfo {author} {\bibfnamefont {D.}~\bibnamefont {Gu\'ery-Odelin}}, \bibinfo {author} {\bibfnamefont {P.}~\bibnamefont {Desbiolles}}, \bibinfo {author} {\bibfnamefont {G.}~\bibnamefont {Ferrari}},\ and\ \bibinfo {author} {\bibfnamefont {J.}~\bibnamefont {Dalibard}},\ }\bibfield  {title} {\bibinfo {title} {Giant spin relaxation of an ultracold cesium gas},\ }\href {https://doi.org/10.1103/PhysRevLett.80.1869} {\bibfield  {journal} {\bibinfo  {journal} {Phys. Rev. Lett.}\ }\textbf {\bibinfo {volume} {80}},\ \bibinfo {pages} {1869} (\bibinfo {year} {1998})}\BibitemShut {NoStop}%
\bibitem [{\citenamefont {Chin}\ \emph {et~al.}(2000)\citenamefont {Chin}, \citenamefont {Vuleti\ifmmode~\acute{c}\else \'{c}\fi{}}, \citenamefont {Kerman},\ and\ \citenamefont {Chu}}]{chin2000high}%
  \BibitemOpen
  \bibfield  {author} {\bibinfo {author} {\bibfnamefont {C.}~\bibnamefont {Chin}}, \bibinfo {author} {\bibfnamefont {V.}~\bibnamefont {Vuleti\ifmmode~\acute{c}\else \'{c}\fi{}}}, \bibinfo {author} {\bibfnamefont {A.~J.}\ \bibnamefont {Kerman}},\ and\ \bibinfo {author} {\bibfnamefont {S.}~\bibnamefont {Chu}},\ }\bibfield  {title} {\bibinfo {title} {High resolution feshbach spectroscopy of cesium},\ }\href {https://doi.org/10.1103/PhysRevLett.85.2717} {\bibfield  {journal} {\bibinfo  {journal} {Phys. Rev. Lett.}\ }\textbf {\bibinfo {volume} {85}},\ \bibinfo {pages} {2717} (\bibinfo {year} {2000})}\BibitemShut {NoStop}%
\bibitem [{\citenamefont {Rodriguez}\ \emph {et~al.}(2009)\citenamefont {Rodriguez}, \citenamefont {McCauley}, \citenamefont {Joannopoulos},\ and\ \citenamefont {Johnson}}]{2009PRA_Casimir_rodriguez}%
  \BibitemOpen
  \bibfield  {author} {\bibinfo {author} {\bibfnamefont {A.~W.}\ \bibnamefont {Rodriguez}}, \bibinfo {author} {\bibfnamefont {A.~P.}\ \bibnamefont {McCauley}}, \bibinfo {author} {\bibfnamefont {J.~D.}\ \bibnamefont {Joannopoulos}},\ and\ \bibinfo {author} {\bibfnamefont {S.~G.}\ \bibnamefont {Johnson}},\ }\bibfield  {title} {\bibinfo {title} {Casimir forces in the time domain: Theory},\ }\href {https://doi.org/10.1103/PhysRevA.80.012115} {\bibfield  {journal} {\bibinfo  {journal} {Phys. Rev. A}\ }\textbf {\bibinfo {volume} {80}},\ \bibinfo {pages} {012115} (\bibinfo {year} {2009})}\BibitemShut {NoStop}%
\bibitem [{\citenamefont {Hung}\ \emph {et~al.}(2013)\citenamefont {Hung}, \citenamefont {Meenehan}, \citenamefont {Chang}, \citenamefont {Painter},\ and\ \citenamefont {Kimble}}]{2013NJP_1dPCW}%
  \BibitemOpen
  \bibfield  {author} {\bibinfo {author} {\bibfnamefont {C.-L.}\ \bibnamefont {Hung}}, \bibinfo {author} {\bibfnamefont {S.~M.}\ \bibnamefont {Meenehan}}, \bibinfo {author} {\bibfnamefont {D.~E.}\ \bibnamefont {Chang}}, \bibinfo {author} {\bibfnamefont {O.}~\bibnamefont {Painter}},\ and\ \bibinfo {author} {\bibfnamefont {H.~J.}\ \bibnamefont {Kimble}},\ }\bibfield  {title} {\bibinfo {title} {Trapped atoms in one-dimensional photonic crystals},\ }\href {https://doi.org/10.1088/1367-2630/15/8/083026} {\bibfield  {journal} {\bibinfo  {journal} {New Journal of Physics}\ }\textbf {\bibinfo {volume} {15}},\ \bibinfo {pages} {083026} (\bibinfo {year} {2013})}\BibitemShut {NoStop}%
\bibitem [{\citenamefont {Asenjo-Garcia}\ \emph {et~al.}(2017{\natexlab{b}})\citenamefont {Asenjo-Garcia}, \citenamefont {Hood}, \citenamefont {Chang},\ and\ \citenamefont {Kimble}}]{2016PRA_GreenFunction}%
  \BibitemOpen
  \bibfield  {author} {\bibinfo {author} {\bibfnamefont {A.}~\bibnamefont {Asenjo-Garcia}}, \bibinfo {author} {\bibfnamefont {J.~D.}\ \bibnamefont {Hood}}, \bibinfo {author} {\bibfnamefont {D.~E.}\ \bibnamefont {Chang}},\ and\ \bibinfo {author} {\bibfnamefont {H.~J.}\ \bibnamefont {Kimble}},\ }\bibfield  {title} {\bibinfo {title} {Atom-light interactions in quasi-one-dimensional nanostructures: A green's-function perspective},\ }\href {https://doi.org/10.1103/PhysRevA.95.033818} {\bibfield  {journal} {\bibinfo  {journal} {Phys. Rev. A}\ }\textbf {\bibinfo {volume} {95}},\ \bibinfo {pages} {033818} (\bibinfo {year} {2017}{\natexlab{b}})}\BibitemShut {NoStop}%
\bibitem [{\citenamefont {Luiten}\ \emph {et~al.}(1996)\citenamefont {Luiten}, \citenamefont {Reynolds},\ and\ \citenamefont {Walraven}}]{1996PRA_evaporativeCooling}%
  \BibitemOpen
  \bibfield  {author} {\bibinfo {author} {\bibfnamefont {O.~J.}\ \bibnamefont {Luiten}}, \bibinfo {author} {\bibfnamefont {M.~W.}\ \bibnamefont {Reynolds}},\ and\ \bibinfo {author} {\bibfnamefont {J.~T.~M.}\ \bibnamefont {Walraven}},\ }\bibfield  {title} {\bibinfo {title} {Kinetic theory of the evaporative cooling of a trapped gas},\ }\href {https://doi.org/10.1103/PhysRevA.53.381} {\bibfield  {journal} {\bibinfo  {journal} {Phys. Rev. A}\ }\textbf {\bibinfo {volume} {53}},\ \bibinfo {pages} {381} (\bibinfo {year} {1996})}\BibitemShut {NoStop}%
\bibitem [{\citenamefont {Hung}\ \emph {et~al.}(2008)\citenamefont {Hung}, \citenamefont {Zhang}, \citenamefont {Gemelke},\ and\ \citenamefont {Chin}}]{2008PRA_Hung}%
  \BibitemOpen
  \bibfield  {author} {\bibinfo {author} {\bibfnamefont {C.-L.}\ \bibnamefont {Hung}}, \bibinfo {author} {\bibfnamefont {X.}~\bibnamefont {Zhang}}, \bibinfo {author} {\bibfnamefont {N.}~\bibnamefont {Gemelke}},\ and\ \bibinfo {author} {\bibfnamefont {C.}~\bibnamefont {Chin}},\ }\bibfield  {title} {\bibinfo {title} {Accelerating evaporative cooling of atoms into {B}ose-{E}instein condensation in optical traps},\ }\href {https://doi.org/10.1103/PhysRevA.78.011604} {\bibfield  {journal} {\bibinfo  {journal} {Phys. Rev. A}\ }\textbf {\bibinfo {volume} {78}},\ \bibinfo {pages} {011604} (\bibinfo {year} {2008})}\BibitemShut {NoStop}%
\end{thebibliography}%

\end{document}